\documentclass[journal]{IEEEtai}
\usepackage[colorlinks,urlcolor=blue,linkcolor=blue,citecolor=blue]{hyperref}
\usepackage{amsmath,amsfonts}
\usepackage{array}
\usepackage[caption=false,font=normalsize,labelfont=sf,textfont=sf]{subfig}
\usepackage{stfloats}
\usepackage{url}
\usepackage{graphicx}
\usepackage{float}
\usepackage{balance}
\usepackage{bm}
\usepackage{cite}
\usepackage{threeparttable}
\usepackage{footnote}
\usepackage{multirow}
\usepackage{multicol}
\usepackage{makecell}
\usepackage{booktabs}

\def\BibTeX{{\rm B\kern-.05em{\sc i\kern-.025em b}\kern-.08em
    T\kern-.1667em\lower.7ex\hbox{E}\kern-.125emX}}

\makeatletter
\long\def\@makecaption#1#2{\ifx\@captype\@IEEEtablestring%
\footnotesize\begin{center}{\normalfont\footnotesize #1}\\
{\normalfont\footnotesize\scshape #2}\end{center}%
\@IEEEtablecaptionsepspace
\else
\@IEEEfigurecaptionsepspace
\setbox\@tempboxa\hbox{\normalfont\footnotesize {#1.}~~ #2}%
\ifdim \wd\@tempboxa >\hsize%
\setbox\@tempboxa\hbox{\normalfont\footnotesize {#1.}~~ }%
\parbox[t]{\hsize}{\normalfont\footnotesize \noindent\unhbox\@tempboxa#2}%
\else
\hbox to\hsize{\normalfont\footnotesize\hfil\box\@tempboxa\hfil}\fi\fi}
\makeatother

\newcommand{\norm}[1]{\lVert#1\rVert}

\begin{document}
\title{Inferring Electrocardiography From Optical Sensing Using Lightweight Neural Network}
\author{Yuenan Li, \IEEEmembership{Senior Member, IEEE}, Xin Tian, \IEEEmembership{Member, IEEE}, Qiang Zhu, \IEEEmembership{Member, IEEE},\\and Min Wu, \IEEEmembership{Fellow, IEEE}
\thanks{\textcopyright 2024 IEEE. Personal use of this material is permitted.  Permission from IEEE must be obtained for all other uses, in any current or future media, including reprinting/republishing this material for advertising or promotional purposes, creating new collective works, for resale or redistribution to servers or lists, or reuse of any copyrighted component of this work in other works.}
\thanks{This work was supported in part by NSF under Grant 2124291. (\textit{Corresponding author: Min Wu.})}
\thanks{Yuenan Li is with the School of Electrical and Information Engineering, Tianjin University, Tianjin 300072, China (email: ynli@tju.edu.cn). This work was initiated when Yuenan Li was a visiting researcher at the University of Maryland, College Park. }
\thanks{Xin Tian and Qiang Zhu were with the Department of Electrical and Computer Engineering, University of Maryland, College Park, MD 20742 USA, where the work was carried out, and are now with Meta Inc., Menlo Park, CA 94025 USA (e-mail: xtian17@terpmail.umd.edu, zhuqiang@terpmail.umd.edu).}
\thanks{Min Wu is with the Department of Electrical
and Computer Engineering and the Institute for Advanced Computer Studies, University of Maryland, College Park, MD 20742 USA (e-mail: minwu@umd.edu).}
}

\markboth{Accepted by IEEE Transactions on Artificial Intelligence, 2024}
{Li \MakeLowercase{\textit{et al.}}: Inferring Electrocardiography From Optical Sensing Using Lightweight Neural Network}

\maketitle

\begin{abstract}
This paper presents a computational solution that enables continuous cardiac monitoring through cross-modality inference of electrocardiogram (ECG). While some smartwatches now allow users to obtain a 30-second ECG test by tapping a built-in bio-sensor, these short-term ECG tests often miss intermittent and asymptomatic abnormalities of cardiac functions. It is also infeasible to expect persistently active user participation for long-term continuous cardiac monitoring in order to capture these and other types of cardiac abnormalities. To alleviate the need for continuous user attention and active participation, we design a lightweight neural network that infers ECG from the photoplethysmogram (PPG) signal sensed at the skin surface by a wearable optical sensor. We also develop a diagnosis-oriented training strategy to enable the neural network to capture the pathological features of ECG, aiming to increase the utility of reconstructed ECG signals for screening cardiovascular diseases (CVDs). We also leverage model interpretation to obtain insights from data-driven models, for example, to reveal some associations between CVDs and ECG/PPG and to demonstrate how the neural network copes with motion artifacts in the ambulatory application. The experimental results on three datasets demonstrate the feasibility of inferring ECG from PPG, achieving a high fidelity of ECG reconstruction with only about 40K parameters.
\end{abstract}

\begin{IEEEImpStatement}
Existing wearable ECG monitors require users to wear adhesive patches or tap a designated area of a smartwatch during sensing, which is uncomfortable or infeasible for long-term cardiac monitoring. As optical sensors have become ubiquitous in wearable devices, this paper proposes a lightweight neural network for inferring ECG from the optical measure of the pulse wave induced by heartbeat, enabling a wearable device to function as an ECG monitor. This work can support long-term ECG monitoring without having users suffer from adhesive-related skin irritation associated with multi-day recorder patch, or constantly hold on a sensor. The learned model has the potential to build the physiological digital twin that facilitates personalized cardiovascular healthcare. The exploration of model interpretation reveals the influence of heart diseases on physiological signals, and the insights obtained through the proposed mechanism help identify diagnostic markers.
\end{IEEEImpStatement}



\section{Introduction}
\IEEEPARstart{C}{ardiovascular} diseases (CVDs) are the most prevalent causes of mortality. According to the statistics in a study conducted in 2019 \cite{CDC}, one person dies from CVDs every 37 seconds in the United States. Early treatment can effectively reduce the risk of sudden cardiac death. However, some CVDs, such as heart muscle dysfunction, show no obvious symptoms in the early stage. The presence of symptoms usually indicates the onset of heart failure. A study conducted in the aged population shows that around one-third to one-half of heart attacks are clinically unrecognized \cite{Over55}. The unawareness of diseases makes some patients lose the opportunity to receive early medical intervention.

Electrocardiogram (ECG) is an essential tool for non-invasive diagnosis of CVDs. The patients at higher risk, such as the aging population, can benefit from ECG monitoring. Among the currently available options for continuous ECG monitoring, the Holter monitor is bulky to wear; newer devices attached to the chest with adhesives, such as the Zio Patch, are lightweight, but the prolonged use of adhesives with multi-day monitoring may increase the risk for skin irritations, especially for persons with sensitive skins. These patch-type sensors may slide or fall off under excessive sweating, or restrict patterns from certain activities. Recent technical advances have integrated bio-sensors into smart wearable devices. For example, using the crown and back crystal as electrodes, an Apple Watch allows its user to take ECG tests for up to 30 seconds at a time from the wrist by tapping the crown. Still, asymptomatic and intermittent events could be missed, and continuous user participation by keeping his/her hand on the sensor is impractical. It is desirable that smart wearable devices can continuously monitor cardiac conditions for extended periods without requiring users' active participation.

Researchers have been working toward the above objective through cross-modality signal inference. The pilot study of Zhu \textit{et al.} explored the possibility of inferring ECG from photoplethysmography (PPG) \cite{DCT,DCT-IoT}. PPG manifests the variation of blood volume caused by the movements of the heart muscle. The signal can be sensed by an optical sensor attached to the wrist or finger, without a user consciously participating all the time. Since PPG carries useful vital signs, the miniaturized PPG sensor has become an integral part of smart wearable devices. Using artificial intelligence (AI) and machine learning (ML) algorithms to augment PPG sensing data for ECG monitoring can alleviate the need to re-design bio-sensor hardware and mitigate continuous user attention to carry out the sensing. The software Apps can be seamlessly integrated into existing devices. In light of the advantages of optical sensing in convenience, availability, and cost, some novel PPG-based biomedical schemes have been developed to complement the conventional ECG-based ones, aiming to utilize cellphones or smartwatches to monitor cardiac health at a larger scale. For example, the Apple Heart Study uses smartwatches to detect atrial fibrillation from intermittent wrist PPG \cite{Apple-Heart}.

Exploiting PPG sensors for long-term ECG monitoring also enables in-home health management and expands the geographic reach of cardiologists. Effective techniques for reconstructing ECG from PPG have the potential to provide general users with the opportunities of long-term ECG monitoring (e.g., for weeks or months long), and those with chronic cardiovascular problems and lifetime risk-assessment needs are among the primary beneficiaries. This class of techniques can enable the creation of the cardio-physiological digital twin, helping cardiologists dynamically monitor the status of a patient's heart over time and provide timely and targeted medical assistance \cite{bruynseels2018digital}.

More specifically, ECG and PPG are two sensing modalities of the same physiological process recorded by electrical and optical measures, respectively. Their correlation establishes the feasibility of estimating one from the other. Previous studies validate that the vital signs derived from PPG and ECG show strong agreement \cite{PPG-ECG-Corr}. As will be seen in this paper, our analytical results show that the causal influence of the heart on blood circulation can be clearly observed from PPG and ECG waveforms. In this work, we leverage deep learning to infer ECG from PPG, aiming to achieve low-cost, user-friendly, and interpretable continuous cardiac monitoring. The main contributions of this work are summarized as follows:

1) We propose a lightweight neural network for deriving ECG from PPG. The network captures the correlations between ECG and PPG at multiple scales by taking advantage of the interactions among the convolutional layers with different receptive fields. For faithful representations of pathological ECG patterns, we use a diagnosis-oriented training algorithm to regularize the distribution of reconstructed ECG signals in a feature space that is discriminative of CVDs.

2) Considering the resource-constrained nature of wearable devices, we propose a model compression algorithm to further lower the memory consumption and computational complexity of ECG reconstruction. The knowledge learned by the original network is transferred to a compressed one via attention-based knowledge distillation. To the best of our knowledge, this is the first attempt at using a lightweight neural network to enable the cross-modality inference of ECG.

3) Beyond algorithm design, we focus on explainability. We use gradient-based interpretation to check if the CVD-related features of ECG learned from data for regularizing ECG reconstruction are clinically plausible. This work also addresses the ambulatory application, and the influence of motion on PPG-based ECG inference is examined using causal analysis. Based on that, we take advantage of the motion information sensed by wearable devices to enhance the robustness of ECG reconstruction during exercise, and the effects of the auxiliary information are analyzed via model interpretation.

The rest of this paper is organized as follows. Section \ref{sec:related} briefly reviews related work. Section
\ref{sec:methods} elaborates on the neural network architecture, training, and model compression algorithms. Experimental results and discussions are presented in Section \ref{sec:experiments}. Finally, Section \ref{sec:conclusion} presents conclusions.

\section{Related Work}
\label{sec:related}

The research on PPG-based ECG inference is still in its infancy. A few prior studies have been dedicated to this problem \cite{DCT,DCT-IoT,SC-PPG2ECG,SCLC-PPG2ECG,TANN,WGAN-P2E,W-Net,Bi-LSTM,Performer}. Going beyond the previous capability of mainly estimating the parameters of ECG from PPG \cite{PPG2ECGPara}, Zhu \emph{et al.}'s work first presents an interpretable framework based on signals and systems theories and demonstrates the feasibility of generating ECG waveforms from PPG sensor via computational approach \cite{DCT}. This work translates PPG to ECG in the Discrete Cosine Transform (DCT) domain using linear regression. Tian \emph{et al.} cast PPG-to-ECG mapping as a cross-domain sparse coding problem \cite{SC-PPG2ECG}. The algorithm learns PPG and ECG dictionaries and a linear transform for domain translation simultaneously. Reference \cite{SCLC-PPG2ECG} presents a more discriminative variant of the sparse-coding-based algorithm, where dictionary learning is regularized by CVD labels to enhance the sensitivity of the sparse codes to pathological ECG patterns. A heartbeat comprises a sequence of short-term actions that determine the local morphologies of PPG and ECG. Hence, hand-crafted models have limited flexibility in capturing the multi-scale correlation between PPG and ECG due to using linear mapping or global bases in waveform analysis and synthesis.

Some recent studies leverage the strong expressive power of neural networks to enhance the accuracy of ECG reconstruction. Chiu \emph{et al.} proposed a neural-network-based end-to-end algorithm \cite{TANN}. The algorithm first uses transformation-and-attention networks to modulate raw PPG and then uses a pair of encoder and decoder to synthesize ECG. The training scheme emphasizes the QRS complex using a region-of-interest (ROI) enhanced loss. An adversarial learning framework was designed to learn the PPG-to-ECG mapping \cite{WGAN-P2E}, where the ECG generator attempts to deceive a discriminator by making synthesized waveforms indistinguishable from the realistic ones. Tang \emph{et al.} constructed a W-Net to infer ECG from PPG by cascading double U-Nets \cite{W-Net}. Bi-directional long short-term memory networks (LSTMs) were also adopted to analyze the temporal features of PPG and extract the cues for reconstructing ECG \cite{Bi-LSTM}. Lan's work uses a Transformer-based architecture to analyze the long-range dependencies between PPG and ECG \cite{Performer}, and the reconstructed ECG is combined with PPG for CVD diagnosis. Recently, we have also seen growing research interest in inferring ECG from other sensing modalities. For instance, it has been shown that ECG can be estimated from ballistocardiogram (BCG), which is a micro-signal capturing the slight body motion resulting from the abrupt ejection of blood during heartbeats. Deep learning algorithms were developed to reconstruct ECG from the body surface displacement measured by a wearable gyroscope or continuous-wave radar \cite{AUNet,VibCardiogram,Radar-ECG}.

Neural networks are more effective than hand-crafted models in detecting the traces of short-term heart activities from waveforms, enabling the ECG reconstruction at a higher fidelity. However, deep neural networks often have high demands on storage and computation resources, which hinders mobile applications. Most prior efforts prioritized the fidelity of reconstructed ECG and have not given sufficient attention to the utility of ECG inference for CVD diagnosis. In particular, the explainability of deep-learning-based PPG-to-ECG inference remains unexplored. Existing studies do not elucidate the relationships between PPG and CVDs, while such knowledge is crucial for understanding the impact of CVDs on heart function. The above-mentioned issues motivate us to enhance the model compactness, efficiency, fidelity, and explainability of ECG inference while benefiting downstream clinical applications.

In addition to PPG-based ECG inference, deep learning has been extensively adopted in cardiac signal processing, such as automated PPG and ECG interpretation \cite{Arrhythmia-NM,LocalVent,AggCardiac,Hypertension}, artifacts removal \cite{DenoisingECG}, waveform synthesis \cite{PGANs}, vital sign measurement \cite{BloodPres,PPG-HR}, heart function assessment \cite{Cardic-Indices}, and biometric recognition \cite{ECG-ID}. Hannun \textit{at al.} trained a deep neural network for detecting rhythm-based disorder \cite{Arrhythmia-NM}. The network can classify 12 kinds of arrhythmia from single-lead ECG at cardiologist-level accuracy. Deep learning was also applied to monitor the aggravation of heart diseases from ECG \cite{AggCardiac}. To improve the accuracy of patient-specific CVD diagnosis, Golany \textit{et al.} developed a generative adversarial network (GAN) for synthesizing the ECG waveforms of a patient \cite{PGANs}. Deep learning also facilitates the measurement of vital signs. It has been demonstrated that blood pressure may be inferred from PPG using a deep belief network \cite{BloodPres}, making it promising to monitor continuous blood pressure in a cuffless way.

\section{Methods}
\label{sec:methods}

\subsection{Physiological Background and Problem Formulation}
\label{sec:background}

ECG measures the electrical signal generated by the depolarization and re-polarization of heart muscle cells. These activities are triggered by an electrical stimulus originating from the SA node (i.e., the pacemaker of the heart). The stimulus coordinates the contracting and expanding movements of the heart, which are the driving force of blood circulation.
\begin{figure}[!ht]
\begin{center}
\includegraphics[width=0.95\linewidth]{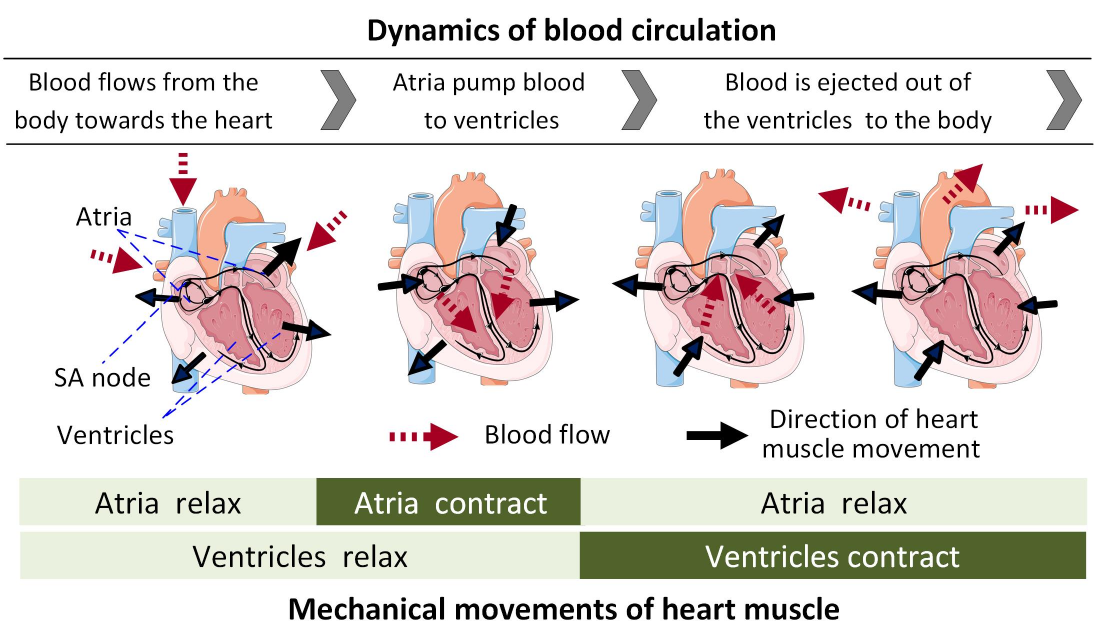}
\caption{Phases of a cardiac cycle.}
\label{Cardiac}
\end{center}
\vspace{-0.4cm}
\end{figure}

\begin{figure}[!h]
\begin{center}
\includegraphics[width=0.95\linewidth]{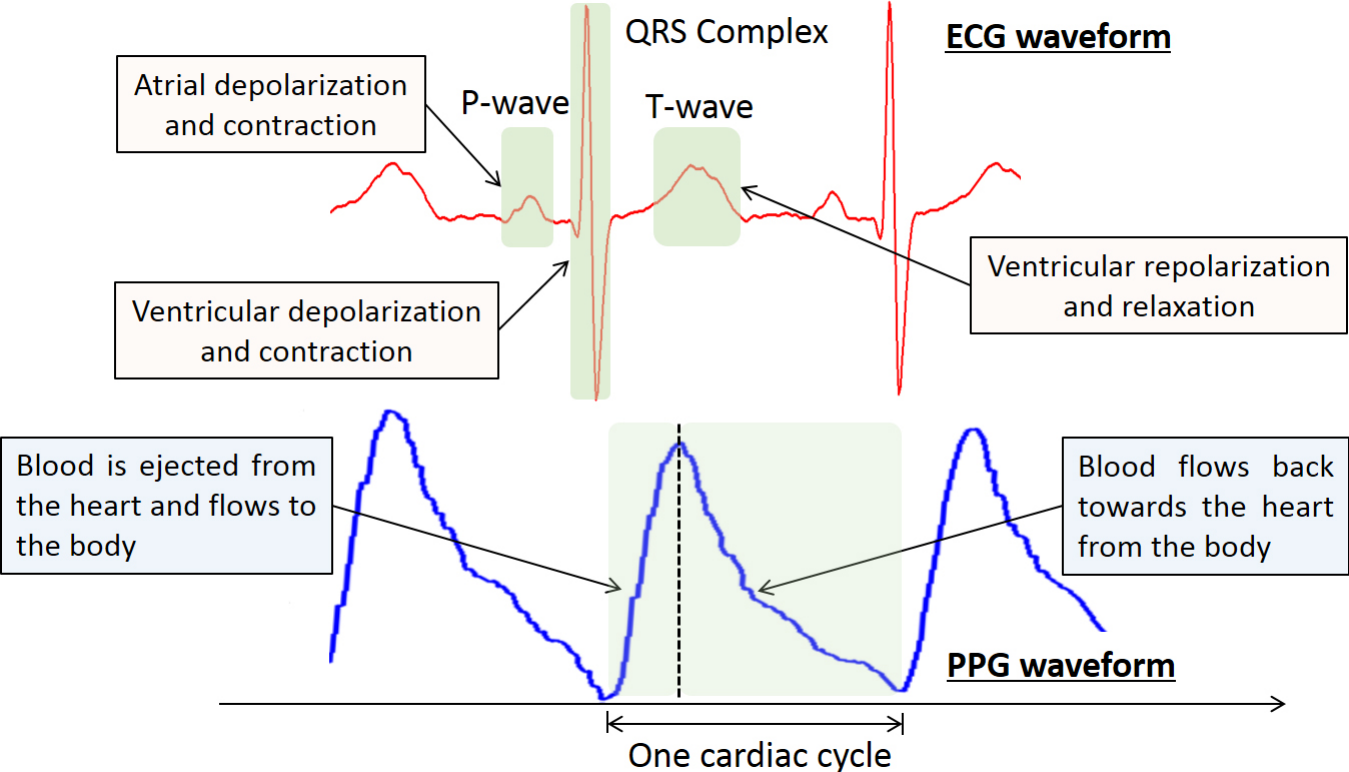}
\caption{Association between the electrical activities of the heart and the blood flow dynamics represented by ECG and PPG.}
\label{ECG-PPG}
\end{center}
\vspace{-0.4cm}
\end{figure}

\begin{figure*}[!t]
\begin{center}
\includegraphics[width=0.98\linewidth]{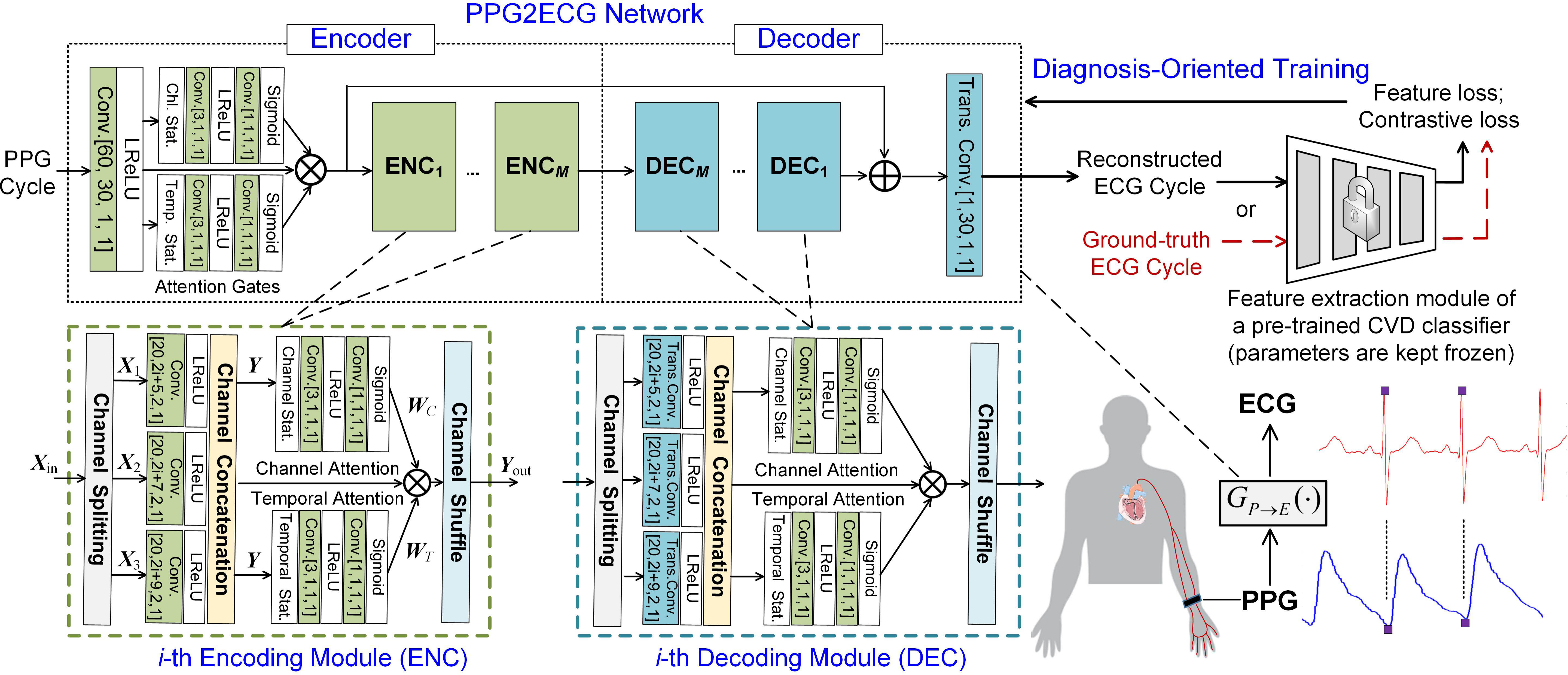}
\caption{Architecture of the PPG2ECG network and the diagnosis-oriented training scheme. Parameter settings of the network are shown in the figure. The dashed boxes below the PPG2ECG network show the architectures of ENCs and DECs. The numbers of ENCs and DECs were set to $M=4$. In each ENC or DEC, input feature maps are split into $G=3$ groups. We use the notation $[N, K, D, S]$ to represent the parameter setting of a 1D convolutional (or transposed convolutional) layer, where $N$ is the kernel number, $K$ is the kernel length, $D$ is the dilation, and $S$ is the stride.}
\label{Arch}
\end{center}
\vspace{-0.5cm}
\end{figure*}

Fig.~\ref{Cardiac} shows the major events during a cardiac cycle\footnote{The heart images in Fig.~\ref{Cardiac} and Fig.~\ref{Arch} are adopted from Servier Medical Art at \url{https://smart.servier.com}, licensed under a Creative Commons Attribution 3.0 unported license.}. A cardiac cycle begins when the four chambers of the heart relax. The two upper chambers (i.e., atria) expand to receive blood from the body. The stimulus first triggers the depolarization of the atria, resulting in the P-wave on ECG. The depolarization causes the atria muscle to contract and pump blood into the two bottom chambers (i.e., ventricles). The electrical stimulus then transmits to the ventricles through the conducting pathway and generates the QRS complex on ECG. As the ventricles contract, blood is ejected out of the heart. More specifically, the left ventricle pumps blood to vessels of the body. The increase of blood volume in the vessels gives rise to an ascending slope on PPG. After that, the ventricles start to relax, and the T-wave on ECG depicts this phase. Finally, both the atria and ventricles relax, so the pressure within the heart drops rapidly, and a new cycle is about to start. As a result, blood flows back toward the atria, leaving a descending slope on PPG. Fig.~\ref{ECG-PPG} depicts the traces of several key cardiac events on ECG and the corresponding blood circulation reflected by PPG.

Consider a simplified conceptual model of the ECG $(\bm{E})$ and PPG $(\bm{P})$ signals. Denoting by $\bm{\delta}$ the electrical stimulus that initiates a heartbeat, we have:
\begin{align}
\label{Sparsity}
\bm{E}=& H_E(\bm{\delta})+\bm{N}_E, \\
\bm{P}=& H_P(\bm{\delta})+\bm{N}_P,
\end{align}

\noindent where $H_E(\cdot)$ and $H_P(\cdot)$ are the impulse response functions describing the electrical and mechanical activations of the heart, respectively, and $\bm{N}_E$ and $\bm{N}_P$ are sensing noise. Without direct access to $H_E(\cdot)$ and $H_P(\cdot)$, inferring ECG from PPG is an under-determined inverse problem. In this work, we solve this problem by training a lightweight neural network $G_{P\rightarrow E}(\cdot)$. Instead of resorting to a highly simplified model with a universal basis of DCT and linear mapping in the prior art \cite{DCT}, we harness data in capturing potentially complex relations while striving to keep the model compact and explainable.

\subsection{Signal Preprocessing}
\label{sec:SP}

The training ECG and PPG sequences are preprocessed using the same procedures as the prior work of Zhu \textit{et al.} \cite{DCT, DCT-IoT} and Tian \textit{et al.} \cite{SC-PPG2ECG, SCLC-PPG2ECG}. We take the moment when the ventricles contract as the anchor point for PPG-ECG synchronization, where the onset points of PPG are aligned to the R-peaks of ECG. The detrending algorithm adopted by Zhu \textit{et al.} \cite{DCT} is then applied on aligned sequences to eliminate the slow-varying trends introduced by breathing, motion, etc. The detrended sequences are partitioned into cycles. Each cycle starts at an onset point of PPG or an R-peak of ECG. The PPG and ECG cycles are then interpolated to fixed length as $\bm{P}\in \mathbb{R}^L$ and $\bm{E} \in \mathbb{R}^L$, respectively.

\subsection{Neural Network Architecture}
\label{sec:network}

At the core of PPG-based ECG reconstruction (a.k.a. PPG2ECG) is mining the intrinsic correlation between PPG and ECG. The cardiac events within a heartbeat are of different durations, and the electro-mechanical activities of the heart are multi-scale in nature \cite{Cardic-Model}. For example, the contraction of the atria lasts about 0.1s, while the blood ejection phase of the ventricles has a much longer duration of about 0.4s. To effectively capture their representations on PPG and ECG, the neural network needs to explore the signal space at various scales. Hence, a key challenge in designing a lightweight PPG2ECG model is maintaining a strong capability of multi-scale feature learning/fusion under a strict constraint on parameter budget.

The proposed PPG2ECG network transforms PPG to ECG using an encoder-decoder architecture. As Fig.~\ref{Arch} shows, the network consists of a stack of multi-scale feature extraction and waveform reconstruction modules, and the building blocks are referred to as encoding and decoding modules. It is worth noting that such an encoding-and-decoding framework, also known as an analysis-and-synthesis framework, is common to many representation and inference tasks in learning as well as signal processing. Using neural networks to  accomplish the encoding/analysis and decoding/synthesis offers flexibility beyond simple models (such as linear and/or parametric ones). The encoder progressively aggregates input PPG to latent codes. Each module characterizes the local morphologies of PPG at multiple temporal resolutions, aiming to detect the short-term and long-term influence of heart activities on blood circulation. The decoder synthesizes ECG using the latent codes from coarse to fine scales. The encoder first uses a 1D convolutional layer and attention gates to extract the primary features of PPG. A stack of encoding modules (ENC) then extracts from PPG the information that is useful for synthesizing ECG. In each ENC, the input feature map $\bm{X}_\textrm{in}\in \mathbb{R}^{C\times L_{\textrm{in}}}$ is first split along the channel direction into $G$ non-overlapping groups, as $\{\bm{X}_i \in \mathbb{R}^{\frac{C}{G}\times L_{\textrm{in}}}|i=1,\cdots, G \}$. The groups are fed to $G$ 1D convolutional layers (see the detailed view of ENC in Fig.~\ref{Arch}) whose kernel lengths increase with a step size of 2 from the top-down direction. The outputs of the convolutional layers, which are of the same length, are concatenated along the channel direction:

\begin{equation}
\label{Conv}
\bm{Y}=\left[ C_1(\bm{X}_1), \cdots, C_G(\bm{X}_G) \right]\in \mathbb{R}^{C\times L_{\textrm{out}}},
\end{equation}

\noindent where $C_i(\cdot)$ is the $i$-th convolutional layer, and $\left[,\cdots,\right]$ denotes the concatenation operation. $\bm{Y}$ encodes the temporal characteristics of the input extracted at $G$ different scales.

A major benefit of grouped convolutions is reducing parameters. Take an ENC without channel splitting (i.e., $G=1$) as a reference. Assume that both the input and output feature maps have $C$ channels and the lengths of 1D convolutional kernels are $K$. Then the number of parameters is $P_1=C^2 K$, where we omit the bias terms for simplicity. For an ENC with $G$ groups $(G>1)$, each group takes $C/G$ channels as input and outputs the same number of channels. If we increase the kernel length by two at a time starting at $K$, the total number of parameters becomes

\begin{equation}
\label{Para-Num}
P_G=\sum_{i=1}^G\left[K+2(i-1)\right]\left(\frac{C}{G}\right)^2 = \frac{\left(K+G-1\right)C^2 }{G} \approx \frac{P_1}{G}.
\end{equation}
\noindent We set $G=3$ and $K \geq 7$, so the grouped convolution has about $1/3$ the parameter amount compared to the one without channel splitting.

Two attention gates are placed after the convolutional layers in each ENC to re-weight the feature map $\bm{Y} \in \mathbb{R}^{C\times L_{\textrm{out}}}$ along the temporal and channel directions. Since blood circulation is driven by the instantaneous activation of the heart muscle, the cues for inferring the activities of the heart do not spread uniformly over the PPG cycle. The PPG2ECG network uses a statistics-based temporal attention gate to highlight the parts of PPG that are informative for synthesizing ECG. We compute the statistics of each column of $\bm{Y}$, including the mean value, max value, and standard deviation. Two 1D convolutional layers are applied to the column-wise statistics to generate weights $\bm{W}_T \in [0, 1]^{L_{\textrm{out}}}$. Similarly, the channel attention gate learns to re-weight $\bm{Y}$ based on row-wise statistics. The statistics reflect the distribution of the features extracted by each convolutional kernel, according to which the attention gate assigns a weight to each channel, giving $\bm{W}_C \in [0, 1]^C$. The feature map is modulated as follows:

\begin{equation}
\label{Att}
\bm{Y}_\textrm{out} = \bm{Y} \otimes \bm{W}_T \otimes  \bm{W}_C,
\end{equation}
\noindent where $\otimes$ is the element-wise multiplication in the corresponding dimension. The attention gates are also appended at the end of the first convolutional layer.

\begin{figure}[!ht]
\begin{center}
\includegraphics[width=1\linewidth]{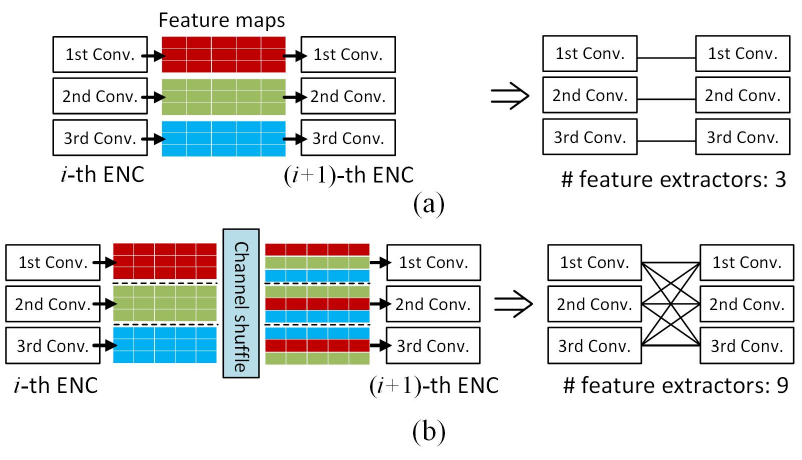}
\vspace{-0.6cm}
\caption{Effect of channel shuffle. (a) and (b) illustrate the cases without and with channel shuffle, respectively, and the figures at the right illustrate the equivalent connection among the convolutional layers in neighboring ENCs.}
\label{Shuffle}
\end{center}
\end{figure}

The modulated feature map is shuffled along the channel direction before being fed to the next ENC (see the detailed view of ENC in Fig.~\ref{Arch}), and we use the transpose-based shuffle operation \cite{ShuffleNet}. The shuffle layer is parameter-free and can increase the richness of features by virtually connecting all the convolutional layers in different ENCs. Fig.~\ref{Shuffle} demonstrates the effect of channel shuffle. As Fig.~\ref{Shuffle}(a) shows, without channel shuffle, the output of a convolutional layer in the first ENC only flows to the corresponding layer in the next ENC. Accordingly, stacking ENCs results in three parallel convolutional pipelines, which is equivalent to independently applying three feature extractors over the input. By contrast, the shuffle operation gives the output of a convolutional layer equal opportunities to flow to all the convolutional layers in the next ENC, as illustrated in the right part of Fig.~\ref{Shuffle}(b). By virtually connecting the convolutional layers in sequential ENCs, the channel splitting and shuffle operations increase the number of pathways for data flow within the neural network, and each pathway serves as a feature extractor. Since the convolutional layers in each ENC are of different kernel sizes, the neural network extracts features from PPG at various scales as the input flows through these pathways, and the characteristics of output ECG can be inferred at multiple granularities as well. As mentioned earlier, the movements of the atria and ventricles within a heartbeat are highly variable in time duration. Wider coverage of scales boosts the feature learning capability of the PPG2ECG neural network, which is beneficial for analyzing the responses of heartbeats on PPG.

The decoding modules (DECs) have the same architecture as ENCs, except that they use transposed convolutional layers to expand input feature maps. The output of the last DEC is fed to a transposed convolutional layer to reconstruct ECG. As the input PPG goes through cascaded ENCs and DECs, the high-resolution timing information about cardiac events carried by the original waveform is attenuated. To compensate, we follow the best practice of deep neural network design, as seen in the ResNet development \cite{ResNet}, and link the first convolutional layer in the encoder and the final ECG reconstruction layer with a residual connection.

\subsection{Diagnosis-Oriented Training Algorithm}
\label{sec:DOT}

Apart from fidelity, the PPG2ECG network also needs to guarantee the effectiveness of reconstructed ECG waveforms in screening CVDs. In automated and manual diagnosis, not all the ECG sample points contribute equally to the decision, and some CVDs cause local abnormalities. For example, myocardial infarction (a heart attack caused by the obstruction of the blood supply to the heart) sometimes manifests as an elevated ST segment. The reconstructed waveform should preserve such diagnostic features, but the norm between reconstructed ECG and the ground truth cannot emphasize the clinically significant parts of ECG. Therefore, it is necessary to use a task-driven loss to regularize ECG generation. In this work, we use the prior knowledge about CVDs learned from data to regularize the PPG2ECG network. We first train a classifier to detect CVDs from ECG, and the intermediate features, which are more sensitive to pathological ECG patterns, are exploited for training the PPG2ECG network. As validated by ablation experiments, the diagnosis-oriented training scheme makes reconstructed ECG waveforms show higher fidelity and more accurate diagnostic results (on both deep-learning-based and conventional CVD classifiers).

\begin{figure}[!ht]
\begin{center}
\includegraphics[width=1\linewidth]{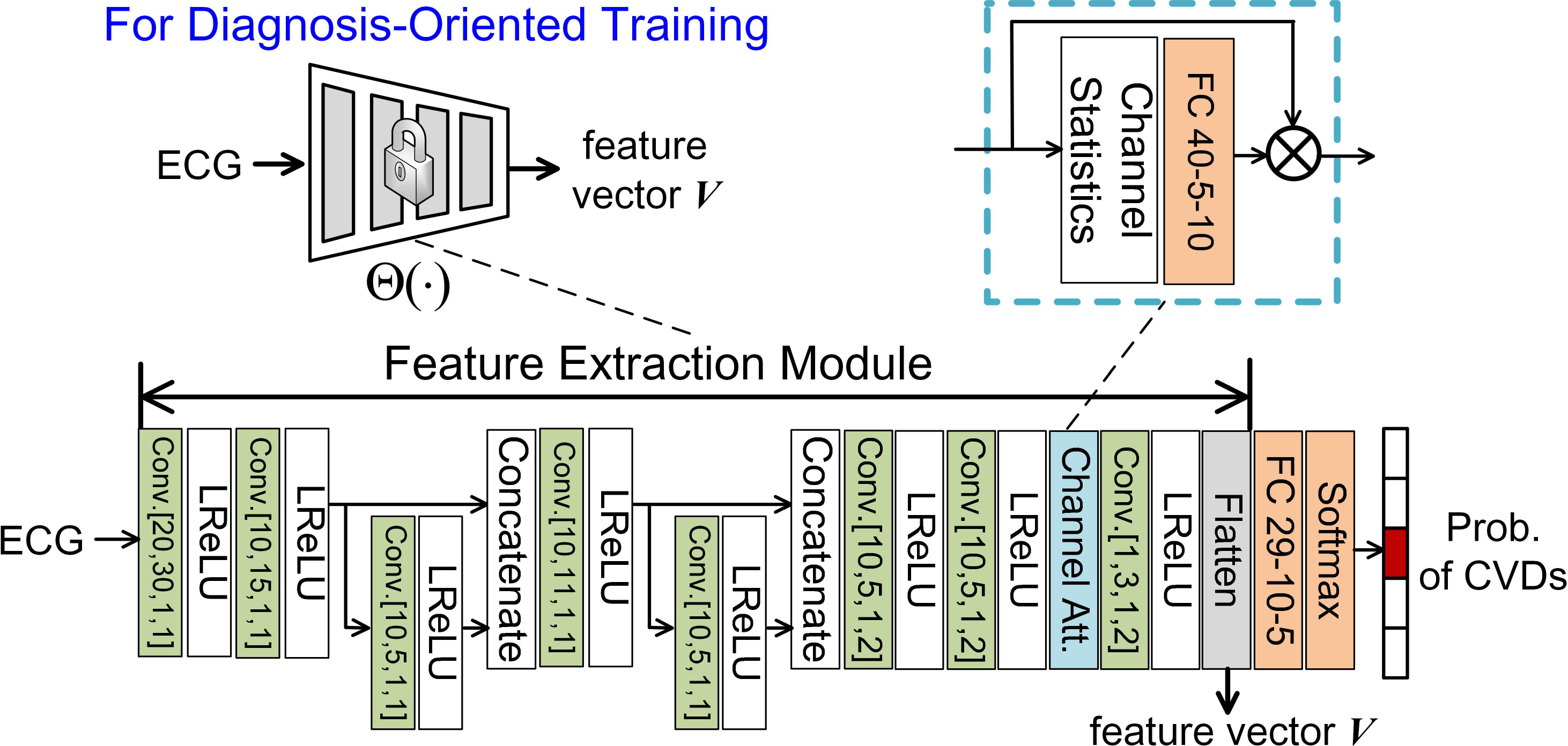}
\vspace{-0.2cm}
\caption{Architecture of the CVD classifier for regularizing ECG inference.}
\label{Classifier}
\end{center}
\end{figure}

The architecture of the classifier is shown in Fig.~\ref{Classifier}. The classifier is composed of cascaded convolutional layers, a squeeze-and-excitation attention gate \cite{SEN}, and fully-connected (FC) layers with softmax output. The features modulated by the attention gate are flattened to a vector and then fed to the fully-connected layers to infer the probability of each disease. The classifier is trained to minimize the cross-entropy loss between the predicted probability vector $\bm{p}$ of CVDs and the one-hot vector $\bm{l}$ of the ground-truth label.

\begin{figure}[!ht]
\begin{center}
\includegraphics[width=0.85\linewidth]{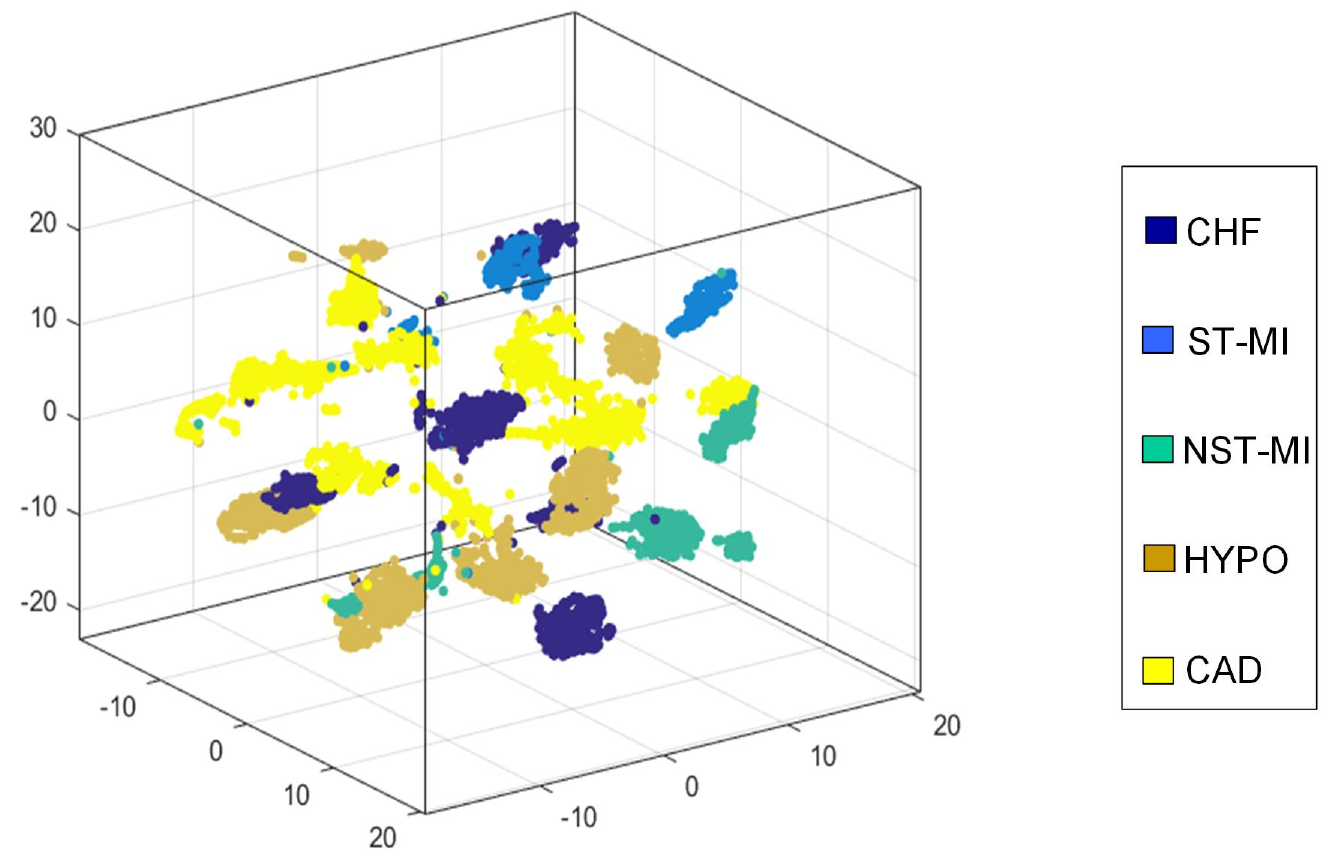}
\caption{Distribution of ECGs in the feature space.}
\label{Clustering}
\end{center}
\vspace{-0.4cm}
\end{figure}

Fig.~\ref{Clustering} visualizes the distribution of the features extracted from the ECGs corresponding to five CVDs, including congestive heart failure (CHF), ST-elevated myocardial infarction (ST-MI), non-ST-elevated myocardial infarction (NST-MI), hypotension (HYPO), and coronary artery disease (CAD). The flattened feature vectors generated by the classifier were plotted in the 3D space using the t-distributed stochastic neighbor embedding (t-SNE) algorithm \cite{t-SNE}, and the classifier was trained using the ECG cycles in the Medical Information Mart for Intensive Care III (MIMIC-III) dataset \cite{MIMIC}. We see that the features are discriminative, and those extracted from the ECGs corresponding to different CVDs are located in distinct clusters.

\begin{figure*}[!ht]
\begin{center}
\includegraphics[width=0.9\linewidth]{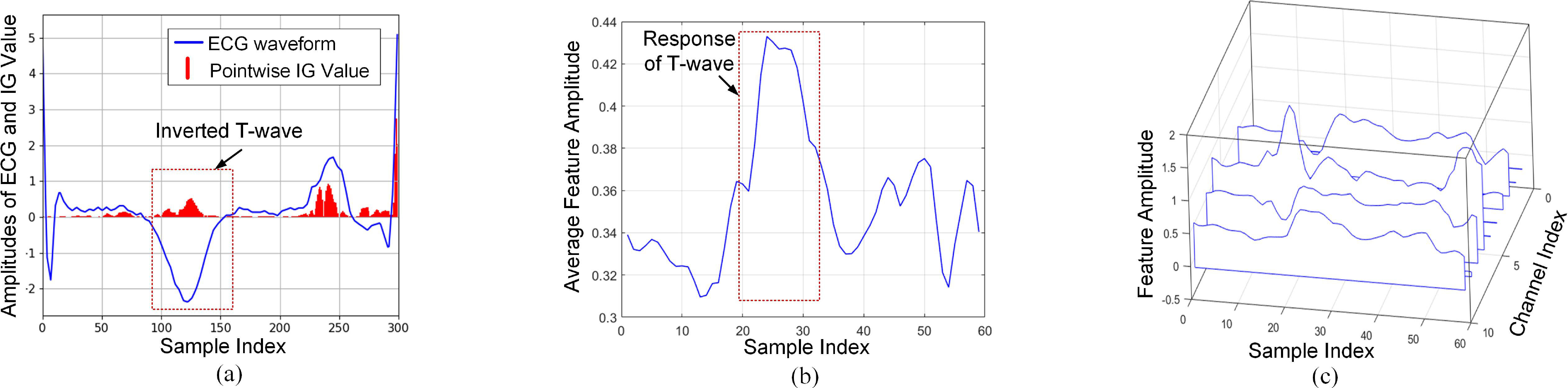}
\caption{Feature visualization and interpretation of the CVD classifier. (a)~An ECG cycle (with inverted T-wave) labeled as NST-MI and the IG values showing the contribution of each sample point to the classification result. (b)~Average feature response of ECG sub-waves. The curve in (b) is plotted by averaging the feature maps in (c) over the channel direction.}
\label{IG-ECG}
\end{center}
\vspace{-0.5cm}
\end{figure*}

Considering the regularization effect of the classifier, it is desirable that the rationale behind its decision and the learned knowledge agree with clinical findings. We interpret the classifier using the axiomatic-attribution-based approach. We adopt the integrated gradient (IG) \cite{IG} method to calculate the contribution of each ECG sample point to the decision made by the classifier. Let us define by $\psi_j(\cdot): \mathbb{R}^L \rightarrow [0, 1]$ the mapping from the input ECG $\bm{E} \in \mathbb{R}^L$ to the $j$-th dimension of the classifier's output (the probability of the $j$-th disease). The IG value of $\bm{E}[i]$ with respect to the disease is:

\begin{equation}
\label{IG-Eq}
\textrm{IG}_i^j= \bm{E}[i] \times \int_{0}^{1} \frac{\partial \psi_j\left(\alpha \bm{E}\right)}{\partial \bm{E}[i]} d\alpha, \quad (i=1,\cdots,L).
\end{equation}

\noindent The amplitude of $\textrm{IG}_i^j$ quantifies the importance of $\bm{E}[i]$ to predicting the $j$-th disease. We use the Riemann sum to approximate the integral.

Fig.~\ref{IG-ECG}(a) plots an ECG cycle labeled as NST-MI and the amplitudes of IG values (represented by red bars). The contributions of ECG sample points to diagnosis are highly uneven, and the classifier pays more attention to several key segments, such as T-wave and P-wave. Note that the ECG cycle has an inverted T-wave, and this abnormal pattern is indicative of ischemia (i.e., decreased blood flow to the heart as a consequence of MI) \cite{ECG-Book}. We also inspected the intermediate features learned by the classifier. The feature maps produced by the attention gate are averaged over the channel direction and plotted in Fig.~\ref{IG-ECG}(b). We observe that the abnormal T-wave shows strong responses in feature maps. The results of model interpretation demonstrate the locality of the clinically significant patterns of ECG and the high sensitivity of the classifier to such patterns. A single per-point distance metric (e.g., the $\ell_1$ norm between reconstructed and target ECG cycles), which gives equal weights to all sample points, cannot emphasize such patterns. To facilitate downstream diagnosis applications, we take advantage of the discriminating power of the classifier. The intermediate features are exploited to help the PPG2ECG network accurately represent the disease-related characteristics of ECG. After training the classifier, we freeze its parameters to define diagnosis-aware training objectives. Denote the feature extraction module of the pre-trained classifier (before the fully-connected layers) by $\Theta(\cdot)$ (see Fig.~\ref{Classifier}). Given a pair of reconstructed ECG cycle $\bm{\hat{E}}$ and the target $\bm{E}$, we compute their feature loss under $\Theta(\cdot)$:

\begin{equation}
\label{Fea-Loss}
\mathcal{L}_F=\norm{\Theta(\bm{E})-\Theta(\bm{\hat{E}})}_2^2.
\end{equation}

Besides pulling $\bm{\hat{E}}$ close to $\bm{E}$ in the feature space, the training algorithm also pushes it far away from those with different CVD labels. As in contrastive learning \cite{Contrast-Learning}, when CVD labels are available, we randomly sample $N$ negative examples $\{\bm{E}_i^-|i=1,\cdots,N\}$ (i.e., the ECG cycles whose labels differ from that of $\bm{E}$) and compute their features using $\Theta(\cdot)$. Let us denote the features of $\bm{\hat{E}}$, $\bm{E}$, and the negative examples as $\bm{\hat{V}}$, $\bm{V}$, and $\{\bm{V}_i^-|i=1,\cdots,N\}$, respectively, and the features are all normalized to unit norm. To prevent class confusion, we use the following contrastive loss to regularize the distribution of ECG cycles in the feature space:
\begin{equation}
\label{Contrast-Loss}
\mathcal{L}_C=\frac{-\textrm{exp}(\bm{\hat{V}} \cdot \bm{V})}{\textrm{exp}(\bm{\hat{V}}\cdot\bm{V})+\sum_{i=1}^N \textrm{exp}(\bm{\hat{V}}\cdot \bm{V}_i^-)},
\end{equation}

\noindent where `$\cdot$' represents the dot product. In summary, the loss function for training the ECG reconstruction network is as follows:
\begin{equation}
\label{All-Loss}
\mathcal{L}=\norm{(\bm{\hat{E}}-\bm{E})\otimes(\bm{1}+\bm{w})}_1+\lambda_F \mathcal{L}_F + \lambda_C \mathcal{L}_C,
\end{equation}

\noindent where $\lambda_F=2$ and $\lambda_C=0.5$ are constant weights. As the training objective designed by Chiu \textit{et al.} \cite{TANN}, we use a weighting vector $\bm{w}\in \mathbb{R}^L$ to enhance the QRS complex, and $\bm{1}$ is an all-ones vector. The weights in $\bm{w}$ are computed using a Gaussian function centered at the R-peak of ECG, and the variance of the Gaussian function is $\sigma^2=1$.

\subsection{Model Compression for Mobile Applications}
\label{sec:Compress}

To better accommodate the stringent memory requirement of mobile devices, we compress the PPG2ECG network using parameter re-use and knowledge distillation.

The cascaded ENCs and DECs take up more than 95\% of the total parameters. The modules have similar architectures but different parameters. If we require the input and output of an arbitrary module to have the same size, the feed-forward pass through $M$ cascaded modules can be simplified by the $R$-depth ($R \leq M$) recursion of one module \cite{Recursive}:

\begin{equation}
\label{Recursion}
\bm{Y_\textrm{out}}= \underbrace{T\circ \cdots \circ T}_R(\bm{X}_\textrm{in}),
\end{equation}

\noindent where $T(\cdot)$ represents the module (ENC or DEC). Take ENC for example, (\ref{Recursion}) is equivalent to repeatedly applying $T(\cdot)$ on the input PPG for $R$ times. In this way, low-level and high-level features are extracted using the same set of kernels. Since the patterns of PPG and ECG are relatively simple, re-using kernels does not noticeably degrade the expressive power of the PPG2ECG network. We have observed in experiments that with the aid of distillation, recursion can reduce over $60\%$ parameters while maintaining the quality of ECG reconstruction. Given a pre-trained PPG2ECG network (i.e., teacher network), we construct a lightweight student network by replacing the cascaded ENCs and DECs with recursive ones, and the student network is then trained to extract knowledge from the teacher network.

\begin{figure}[!ht]
\begin{center}
\includegraphics[width=0.9\linewidth]{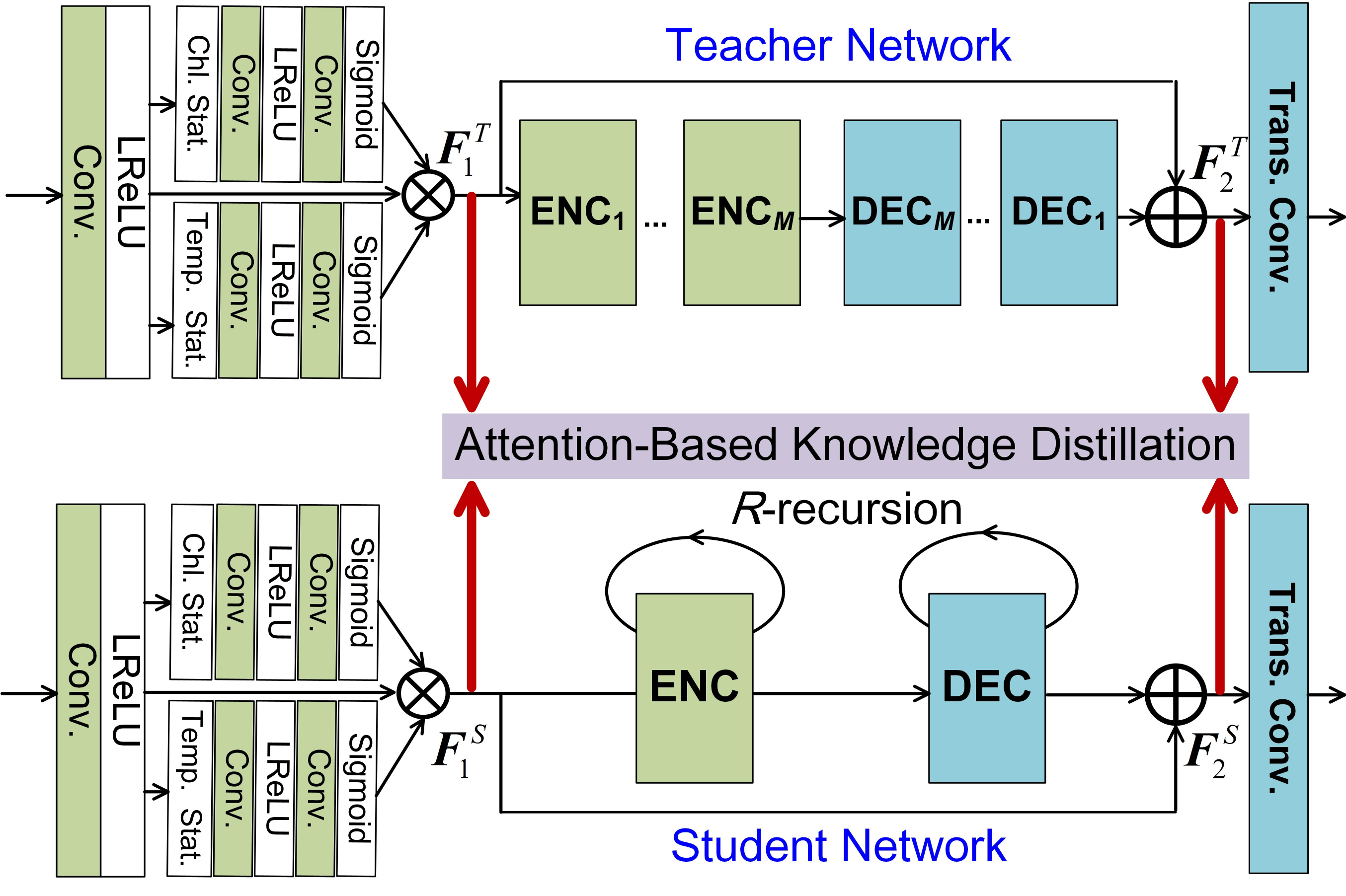}
\caption{Knowledge distillation for training a thin student PPG2ECG network.}
\label{Distill}
\end{center}
\vspace{-0.3cm}
\end{figure}

The student network needs to mimic the internal behaviors of the teacher network to exhibit similar performance. To this end, we select several key positions from the pre-trained teacher network and peek into its internal behaviors from feature maps. For the feature maps generated at each selected position, we apply max-pooling over the channel direction to obtain the maximum feature response at each time instant. The information reflects the salient parts of the feature map and tells where the teacher network pays attention \cite{Att-Distill}. This work selects two positions for knowledge distillation: the inputs to the first ENC and the ECG reconstruction layer (see the red arrows in Fig.~\ref{Distill}). The distillation algorithm forces the student network to reproduce the information distilled from the teacher network while synthesizing ECG. The discrepancy between the two networks is measured by the cosine similarity between the max-pooling results of their intermediate features, leading to the following distillation loss:
\begin{equation}
\label{Distill-Loss}
\mathcal{L}_D= 1-\frac{1}{2} \sum_{i=1}^{2} \frac{\bm{M}_i^T\cdot\bm{M}_i^S}{\norm{\bm{M}_i^T}_2 \norm{\bm{M}_i^S}_2},
\end{equation}

\noindent where $\bm{M}_i^T = \textrm{MaxPool}(\bm{F}^T_i)$ and $\bm{M}_i^S = \textrm{MaxPool}(\bm{F}^S_i)$, $\bm{F}^T_i$ and $\bm{F}^S_i$ are the features sampled from the teacher and student networks at the $i$-th position. When training the student network, we combine $\mathcal{L}_D$ with the fidelity and feature losses in (\ref{All-Loss}), where the ground-truth ECG in (\ref{All-Loss}) is replaced by the output of the teacher network.

\section{Experimental Results and Discussions}
\label{sec:experiments}

\subsection{Datasets, Parameter Settings, and Competing Algorithms}
\label{sec:datasets}

Algorithms were trained and tested on three publicly available datasets. The primary reason for choosing online datasets is to facilitate open benchmarking. The training and testing PPG/ECG waveforms were all sensed from real-world subjects across a broad range of ages, ethnicities, and health conditions, including patients with common CVDs and healthy individuals without any known CVDs. The recording devices include professional monitors in hospitals and consumer-level wearable sensors during physical exercise. The artifacts and sensing noise in raw data ensure comprehensive coverage of real-world scenarios.

\noindent \textbf{MIMIC-III}: MIMIC-III \cite{MIMIC} was chosen as the benchmark dataset for its richness of waveforms and CVD types, detailed diagnostic results from patients, public availability, and the real-world nature. MIMIC-III has a full coverage of pathological patterns related to major CVDs, and noisy data were intentionally preserved to reflect real-world healthcare settings. The waveforms in MIMIC-III allow for evaluating the diagnosis of reconstructed ECG. Per our best knowledge, other publicly available datasets do not have comparable sizes and richness of PPG-ECG patterns and CVD types. Following the practice in previous studies \cite{DCT,SC-PPG2ECG,DCT-IoT,SCLC-PPG2ECG}, we used the waveforms in Folder 35. The waveforms were screened using the signal quality assessment function in the PhysioNet Cardiovascular Signal Toolbox \cite{toolbox}, and those labeled as ``unacceptable (Q)" were discarded.

\noindent \textbf{BIDMC}: The BIDMC PPG and Respiration Dataset was acquired from 53 patients (aged from 19 to 90+) admitted by the Beth Israel Deaconess Medical Centre (BIDMC) \cite{BIDMC}. It provides the 8-min recordings of the PPG, ECG, and physiological parameters (e.g., heart rate and respiratory rate) of each patient. PPG and ECG signals were sampled at a frequency of 125 Hz. The BIDMC and MIMIC-III datasets have no overlap in subjects.

\noindent \textbf{Motion PPG-ECG}: The PPG and ECG waveforms in this dataset were collected during physical exercise \cite{Motion-PPG}. The subjects were asked to run on a treadmill at varying speeds (6km/h, 8km/h, 12km/h, and 15km/h), and there are two rest sessions of 30s at the beginning and end of each recording. The PPG waveforms were measured by a wrist-type sensor with a green LED, and ECG waveforms were sensed by a wet sensor attached to the chest. Moreover, the dataset also provides the acceleration signals simultaneously measured by an accelerometer placed on the wrist. The dataset was created for PPG-based heart rate estimation. The data in the testing part were not used in our experiment since ECG waveforms are not available.

The datasets consist of 71,098 pairs of ECG and PPG cycles, amounting to 47.3 hours of recordings. The signals were normalized to a fixed length of $L=300$. For a fair performance comparison, we followed the same testing protocols as the prior work of Zhu \textit{et al.} \cite{DCT, DCT-IoT} and Tian \textit{et al.} \cite{SC-PPG2ECG, SCLC-PPG2ECG}. The PPG-ECG waveforms of each subject exist in both training and testing datasets. The first 80\% and last 20\% waveforms of each subject were allocated for training and testing, respectively. The parameter settings of the PPG2ECG network are presented in Fig.~\ref{Arch}. The network was implemented in Pytorch and trained for 60 epochs with a batch size of ten using the Adam optimizer (with default parameter settings). The initial learning rate was set to $10^{-3}$ and then decayed by 0.6 every 18 epochs. The proposed algorithm was compared with five pieces of prior work on PPG-based ECG inference, covering both hand-crafted and deep-learning-based approaches, that are the DCT and linear regression algorithm (DCT) \cite{DCT-IoT}, the cross-domain joint dictionary learning algorithm (XDJDL) \cite{SC-PPG2ECG}, the follow-up work that applies label-consistency regularization to XDJDL using CVD labels (LC-XDJDL) \cite{SCLC-PPG2ECG}, the W-shaped neural network (W-Net) \cite{W-Net}, and the transformed attentional neural network (TANN) \cite{TANN}. W-Net was designed for the PPG signals with 1024 samples, so the testing data were first up-sampled to 1024 using bi-linear interpolation, and the reconstructed ECG cycles were then down-sampled to 300 samples for performance evaluation. TANN was tested using the codes posted online. We also adapted and trained a U-Net \cite{U-net} as an additional baseline, and our implementation of the 1D U-Net consists of four convolutional layers and four transposed convolutional layers with 60 kernels each. The kernel lengths of the convolutional layers are 30, 15, 10, and 5, respectively, and stride and dilation were all set to one. Two mirroring convolutional and transposed convolutional layers have the same parameter settings and are linked with a residual connection.

\begin{figure*}[!ht]
\centering
\includegraphics[width=0.85\linewidth]{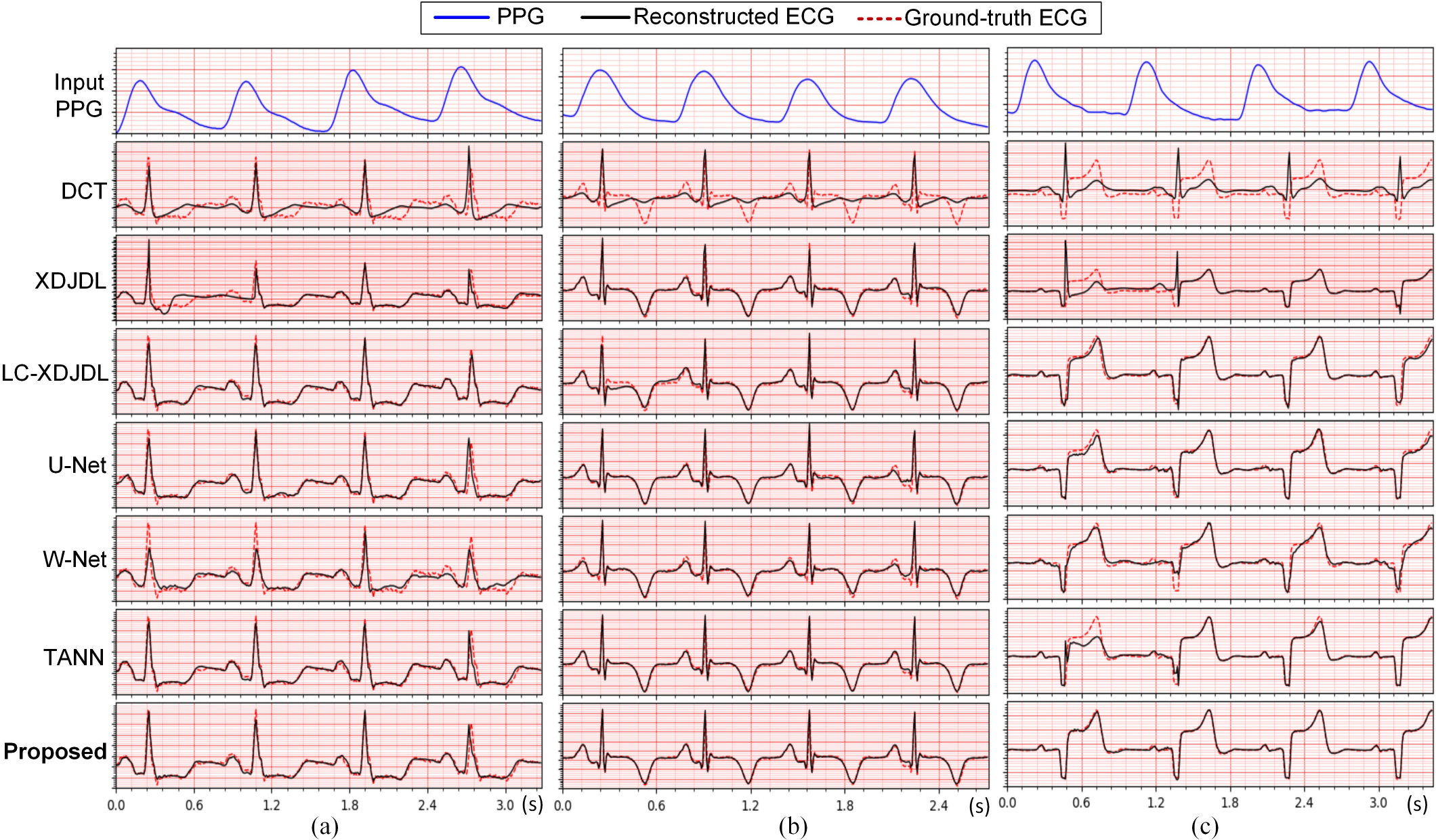}
\caption{Comparison of the ECG sequences inferred by different approaches. The waveforms in (a), (b), and (c) were sensed from the subjects with coronary artery disease, myocardial infarction, and congestive heart failure, respectively.}\label{Recon}
\end{figure*}

\subsection{Comparison on the Quality of ECG Inference, Model Size, Complexity, and Explainability}
\label{sec:Perf}

We use the Pearson correlation coefficient and the relative Root Mean Squared Error (rRMSE) to evaluate the fidelity of inferred ECG $\hat{\bm{E}}$:

\begin{equation}
\label{rho}
\rho = \frac{(\bm{E}-\mu[\bm{E}])^T(\hat{\bm{E}}-\mu[\hat{\bm{E}]})}{\norm{\bm{E}-\mu[\bm{E}]}_2 \; \norm{\hat{\bm{E}}-\mu[\hat{\bm{E}}]}_2},
\end{equation}

\begin{equation}
\label{rRMSE}
\textrm{rRMSE} = \frac{\norm{\bm{E}-\hat{\bm{E}}}_2}{\norm{\bm{E}}_2},
\end{equation}

\noindent where $\bm{E}$ is the ground truth, and $\mu[\cdot]$ represents the element-wise mean value of a vector. Table \ref{QualityScore} lists the statistics of the quality scores measured on each dataset \footnote{LC-XDJDL applies additional label-consistency regularization to XDJDL. It exhibits equal performance as XDJDL on the BIDMC and Motion PPG-ECG datasets where CVD labels are unavailable, so its performance scores on these datasets are not separately listed in Table \ref{QualityScore}.}. Table \ref{Average} gives an overall comparison on average quality scores, parameter amounts, computational complexities, and explainabilities of the testing algorithms. We take the number of floating-point operations (FLOPs) as a platform-independent measure of complexity, and the results in Table \ref{Average} were obtained by counting the operations for reconstructing an ECG cycle with 300 sample points. Two trends can be observed from the comparison. First, compared with the generic orthogonal bases of DCT, data-dependent bases (convolutional kernels and sparse coding atoms) better suit the underlying structures of PPG and ECG. Second, deep-learning-based approaches outperform DCT and XDJDL, which synthesize ECG using global bases. The correlation coefficients and rRMSEs in Table \ref{QualityScore} indicate that the ECG cycles generated by the proposed algorithm have the highest fidelity on all datasets. Our ECG reconstruction model has the least amount of parameters. It also has the highest efficiency among sparse-coding- and deep-learning-based algorithms, and its FLOPs value is nearly half that of U-Net.

\begin{table}[h]
\caption{Quantitative qualities of reconstructed ECGs measured on each testing dataset (Top results are displayed in bold;\newline $\uparrow$ means high value desired, and vice versa.)}\centering\label{QualityScore}
\resizebox{3.5in}{!}{
\begin{tabular}{cccccccc}
\toprule
\multirow{2}{*}{Datasets} & \multirow{2}{*}{Algorithms} & \multicolumn{3}{c}{$\rho$ $\uparrow$} & \multicolumn{3}{c}{rRMSE $\downarrow$}                          \\ \cmidrule(r){3-5} \cmidrule(l){6-8}
 & & $\mu$ &med &$\sigma$ & $\mu$ &med &$\sigma$       \\ \hline
\multirow{5}{*}{MIMIC-III}
 & DCT~\cite{DCT-IoT}   &0.71 & 0.83 &0.31 &0.67 &0.60 &0.26\\
 & XDJDL~\cite{SC-PPG2ECG} &0.88 & 0.96 &0.23 &0.39 &0.29 &0.31\\
 & LC-XDJDL~\cite{SCLC-PPG2ECG} &0.92 & 0.97 &0.17 &0.33 &0.26 &0.25\\
 & U-Net &0.93 & 0.96 &0.13 &0.32 &0.27 &0.19\\
 & W-Net~\cite{W-Net} &0.92 & 0.96 &0.13 &0.33 &0.29 &0.18\\
 & TANN~\cite{TANN} &$\bm{0.94}$ & 0.97 &0.13 &0.28 &0.23 &0.19\\
 & Proposed &$\bm{0.94}$ & $\bm{0.98}$ &0.13 &$\bm{0.27}$ &$\bm{0.22}$ &0.20\\ \hline
 \multirow{5}{*}{BIDMC}
 & DCT~\cite{DCT-IoT} &0.70 & 0.83 &0.35 &0.66 &0.60 &0.23\\
 & XDJDL~\cite{SC-PPG2ECG} &0.82 & 0.94 &0.27 &0.48 &0.35 &0.36\\
 & U-Net &0.86 & 0.93 &0.17 &0.43 &0.37 &0.25\\
 & W-Net~\cite{W-Net} &0.89 & 0.95 &0.14 &0.42 &0.37 &0.22\\
 & TANN~\cite{TANN} &$\bm{0.90}$ & $\bm{0.97}$ &0.17 &$\bm{0.35}$ &$\bm{0.26}$ &0.27\\
 & Proposed &$\bm{0.90}$ & $\bm{0.97}$ &0.16 &$\bm{0.35}$ &$\bm{0.26}$ &0.26\\ \hline
 \multirow{5}{*}{\begin{tabular}[c]{@{}c@{}}\\Motion\\ PPG-ECG\end{tabular}}
 & DCT~\cite{DCT-IoT} &0.55 & 0.79 &0.56 &0.78 &0.69 &0.26\\
 & XDJDL~\cite{SC-PPG2ECG} &0.51 & 0.75 &0.53 &0.79 &0.71 &0.30\\
 & U-Net &0.80 & $\bm{0.83}$ &0.14 &0.58 &$\bm{0.56}$ &0.16\\
 & W-Net~\cite{W-Net} &0.80 & 0.82 &0.12 &0.59 &0.58 &0.13\\
 & TANN~\cite{TANN} &0.80 & $\bm{0.83}$ &0.14 &0.58 &$\bm{0.56}$ &0.17\\
 & Proposed &$\bm{0.81}$ & $\bm{0.83}$ &0.13 &$\bm{0.57}$ &$\bm{0.56}$ &0.16\\
 \bottomrule
 \multicolumn{8}{l}{$\mu$: mean value; med: median value; std: standard deviation.}\\
\end{tabular}}
\end{table}

\begin{table}[h]
\centering
\caption{Comparison on average quality scores, parameter amount, computational complexity, and explainability}\centering\label{Average}
\resizebox{3.5in}{!}{
\begin{tabular}{cccccc}
\toprule
Algorithms & \#Para.(M)$\downarrow$ & FLOPs(M)$\downarrow$ &$\bar{\rho} \uparrow$ &$\overline{\text{rRMSE}} \downarrow$ & Explain.\\\hline
DCT~\cite{DCT-IoT}    & 0.27   & $\bm{0.36}$ & 0.65  & 0.70 &High   \\
XDJDL~\cite{SC-PPG2ECG}  & 5.67   & 60.21 & 0.74  & 0.55  &High \\
LC-XDJDL~\cite{SCLC-PPG2ECG}  & 5.90   & 64.72 & 0.75  & 0.53  &High \\
U-Net  & 0.22   & 55.99 & 0.86  & 0.44  &Low\\
W-Net~\cite{W-Net}  & 2.37   & 35.97 & 0.87  & 0.45  &Low\\
TANN~\cite{TANN}   & 10.98   & 518.79 & $\bm{0.88}$  & $\bm{0.40}$ &Low\\
Proposed   & $\bm{0.12}$   & 28.32 & $\bm{0.88}$  & $\bm{0.40}$  &High \\\bottomrule
\end{tabular}}
\end{table}

The multi-scale and attention mechanisms improve the sensitivity of the proposed network to the subtle difference among PPG waveforms. As can be seen from Fig.~\ref{Recon}, although the waveforms of PPG are quite similar, the network can represent the distinct morphological difference among ECG waveforms. The PPG2ECG network can faithfully infer the fine detail and abnormal morphology of ECG, such as the inverted QRS complex in Fig.~\ref{Recon}(c). The accurate reconstruction of ECG from PPG confirms the intrinsic association between the electrical and mechanical activities of the heart and the existence of CVD-related information in PPG. Besides local morphologies, the timing information of ECG also reflects the health conditions of the heart. For example, a prolonged PR interval indicates slow conduction between the atria and ventricles. We also assessed the accuracy of the PPG2ECG network in representing the timing information. For the waveforms in the MIMIC-III dataset sampled at a frequency of 125Hz, the Mean Absolute Errors (MAE) between the durations of the PR, QRS, and QT intervals measured from reconstructed and ground-truth waveforms are 4ms, 5ms, and 9ms, respectively. The MAE is smaller than a quarter of the horizontal length of a small square (40ms) on ECG graph paper. As can be seen from Fig.~\ref{Recon}, the timing information of the reconstructed and ground-truth waveforms show good agreement. However, it is clear from Table~\ref{QualityScore} that the ECG inference in the ambulatory setting is more challenging. Due to the interference of motion on PPG measurement, all algorithms show degradation on the Motion PPG-ECG dataset. In the following subsection, we will describe a method that exploits auxiliary information for more robust ECG inference.

From the comparison in Table~\ref{Average}, we note that the network exhibits comparable or superior performance than TANN on all the datasets, while its parameter amount and computational load are orders of magnitude lower. It only has about 1\% the parameter amount and 5\% the FLOPs as TANN. The DCT-based algorithm is the most efficient since DCT and linear regression can be implemented through a few matrix-vector multiplications. The computational load required by the proposed algorithm does not impose a substantial burden on mainstream devices. As a reference for comparison, the mobile processor Exynos 7 Dual released in 2016 for wearable devices can perform up to 15G FLOPs per second. We will show later in Section \ref{sec:Exp-Comp} that the model size and computational load can be further reduced via model compression.

We take explainability as another dimension of comparison. Linear models, including DCT \cite{DCT-IoT}, XDJDL \cite{SC-PPG2ECG}, and LC-XDJDL \cite{SCLC-PPG2ECG}, are inherently interpretable. For example, LC-XDJDL uses linear transforms to infer the sparse codes of ECG and CVD labels from the sparse codes of PPG. Hence, the pathological patterns of PPG can be discovered by identifying the atoms that frequently co-occur with a specific CVD in spare representation. Unlike other deep learning algorithms that mainly focus on the data-fitting aspect of PPG-based ECG inference, the proposed work also addresses the interpretation aspect. More specifically, we take PPG-to-ECG mapping and model interpretation as avenues for understanding the influence of CVDs on PPG, as will be discussed later in Section~\ref{sec:CVD-PPG}.

\subsection{Motion-Information-Aided ECG Reconstruction}
\label{sec:Motion-Aided}

Motion artifacts degrade the accuracy of ECG reconstruction. From the physiological perspective, PPG is shaped by the electrical response of the heart. While under intensive physical exercise, motion also interferes with the optical sensing of PPG signal and becomes another causing factor. For validation, we use causal inference to examine the factors that affect PPG, the insights from which help to improve the robustness of ECG reconstruction.

We use directed information \cite{Direct-Info} to analyze the causal influence among PPG, ECG, and motion. Directed information (DI) is an information-theoretic metric for examining the causal influence of one time series on another. Given two $L$-length sequences $\bm{X}$ and $\bm{Y}$, the directed information from $\bm{X}$ to $\bm{Y}$ is denoted by $I(\bm{X} \rightarrow \bm{Y})$, and
\begin{equation}
\label{DI-Def}
I(\bm{X} \rightarrow \bm{Y})=H(\bm{Y})-\sum_{i=1}^{L}H(\bm{Y}[i]\;|\;\bm{Y}[1:i-1],\bm{X}[1:i]),
\end{equation}

\noindent where $H(\cdot)$ represents entropy, $H(\cdot|\cdot)$ represents conditional entropy, and the colon operator has the same meaning as in Matlab.

The DI in the reverse direction is denoted by $I(\bm{Y} \rightarrow \bm{X})$, and the metric is not symmetric. The relationship between the DI values in two directions asserts the causal influence between two series \cite{Direct-Info}:
\begin{enumerate}
\item $I(\bm{X} \rightarrow \bm{Y}) \gg I(\bm{Y}[1:L-1] \rightarrow \bm{X})$ indicates that $\bm{X}$ causes $\bm{Y}$,
\item $I(\bm{X} \rightarrow \bm{Y}) \ll I(\bm{Y}[1:L-1] \rightarrow \bm{X})$ indicates that $\bm{Y}$ causes $\bm{X}$,
\item $I(\bm{X} \rightarrow \bm{Y}) \approx I(\bm{Y}[1:L-1] \rightarrow \bm{X}) \rightarrow 0$ implies the independence of $\bm{X}$ and $\bm{Y}$, and accordingly, there is no causal influence in any direction.
\end{enumerate}

\begin{figure}[!ht]
\begin{center}
\includegraphics[width=1\linewidth]{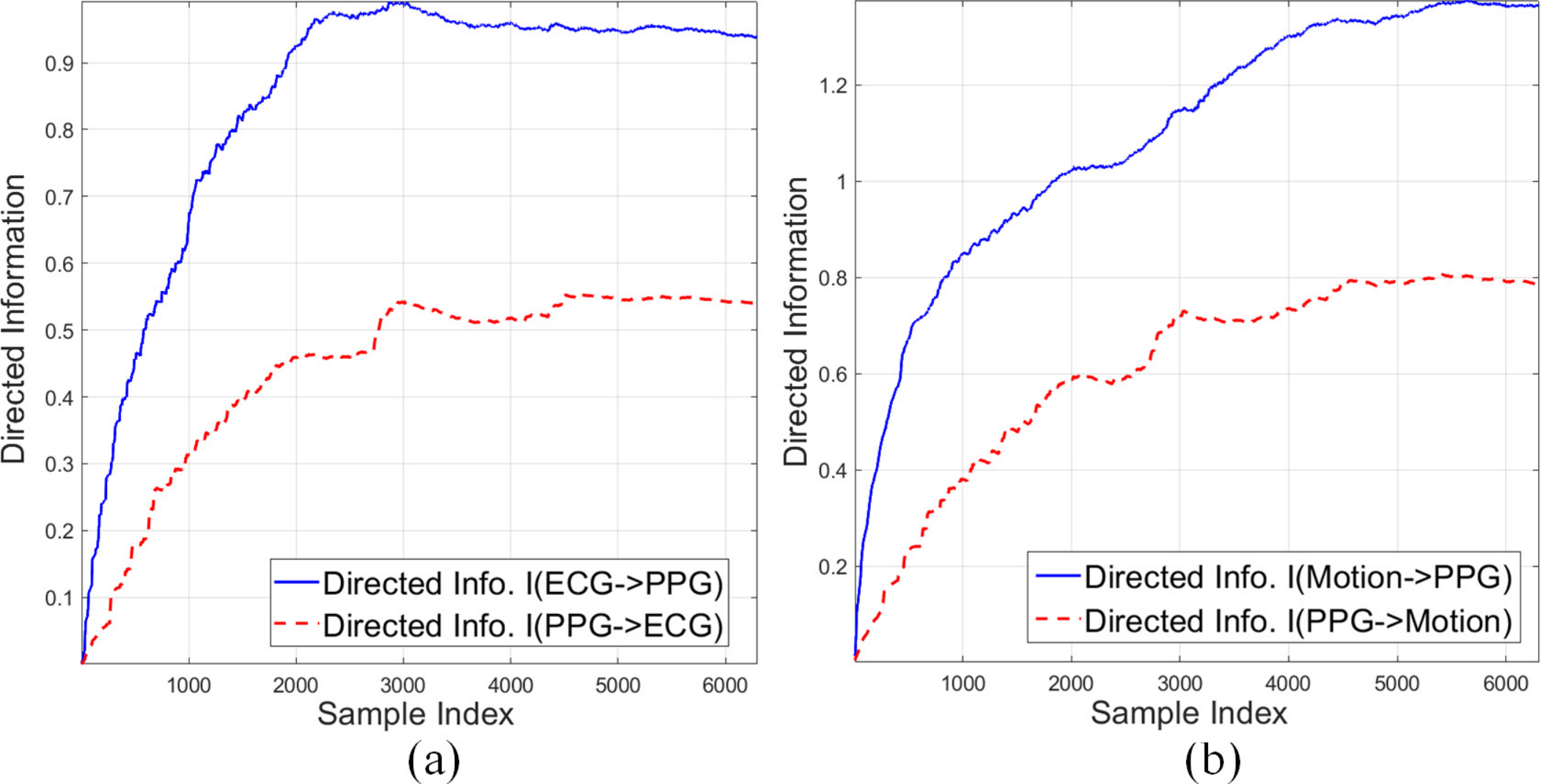}
\vspace{-0.5cm}
\caption{Causal effects of ECG and motion on PPG. (a)~Curves of DI between ECG and PPG. (b)~Curves of DI between acceleration signal and PPG.}
\label{DI-curve}
\end{center}
\vspace{-0.5cm}
\end{figure}

In Fig.~\ref{DI-curve}(a), we plot the curves of the DI values between PPG and ECG waveforms along two directions, and the waveforms were extracted from the Motion PPG-ECG dataset. An obvious causal influence of ECG on PPG can be observed, indicating that the electrical activities of the heart (represented by ECG) cause blood circulation (represented by PPG), which is consistent with known facts. To examine the impact of motion on PPG measurement, we also estimated the DI values between PPG and motion information. The motion information was measured at the subject's wrist using an accelerometer during running. Fig.~\ref{DI-curve}(b) shows the curves of the DI values between PPG and the x-axis acceleration signal. The causal influence of motion on PPG is quite remarkable, suggesting that motion indeed affects the optical sensing of PPG. In this setting, the physiological behavior of the heart is not the sole decisive factor of PPG. Hence, for better robustness of ECG reconstruction, it is necessary to use reference motion information to counteract the artifacts of PPG.

Motion information is readily available since most devices use a built-in accelerometer for fitness tracking. Acceleration signals have been exploited to correct the motion artifacts in PPG. Some PPG-based heart rate estimation algorithms de-noise PPG signals by taking acceleration signal as the reference of adaptive filter \cite{PPG-Acc}. It is usually assumed that motion artifacts are additive in the raw signal space, but the influence of motion on PPG is far more complicated. In this work, we do not directly de-noise PPG. Instead, the three-dimensional acceleration signal is concatenated with PPG along the channel direction as auxiliary inputs. This gives the PG2ECG network more flexibility to alleviate motion artifacts, not only in the raw input space but also in the feature spaces. Table~\ref{Motion} compares the qualities of the ECGs synthesized with and without the auxiliary motion information. The PPG2ECG network aided by acceleration signals achieves more accurate ECG reconstruction, and the average rRMSE drops to 0.51 from 0.57. This finding suggests that using the side information provided by the accelerometer can enhance the robustness of PPG-based cardiac monitoring when a subject performs intensive exercise.

\begin{table}[htbp]
\centering
\caption{Comparison between the performance of ECG reconstruction under exercise with and without the assistance of acceleration signal}\centering\label{Motion}
    \begin{tabular}{ccccccc}
    \toprule
     \multirow{2}{*}{ Modes} & \multicolumn{3}{c}{$\rho$ $\uparrow$} & \multicolumn{3}{c}{rRMSE $\downarrow$}                          \\ \cmidrule(r){2-4} \cmidrule(l){5-7}
  & $\mu$ &med &$\sigma$ & $\mu$ &med &$\sigma$       \\ \hline
      With Acc. Signal   & 0.84 & 0.88 & 0.15 & 0.51 & 0.47 & 0.22   \\
      Without Acc. Signal & 0.81 & 0.83 & 0.13 & 0.57 & 0.56 & 0.16   \\
      \bottomrule
    \end{tabular}
\end{table}

Given the effectiveness of the motion-information-aided scheme, we seek to understand how neural network utilizes motion information in the ambulatory setting using the IG-based model interpretation \cite{IG}. As in interpreting the CVD classifier, we calculated the IG value of the $i$-th dimension of the input with respect to the $j$-th dimension of the output ECG $(\hat{\bm{E}}\in \mathbb{R}^L)$ and denote the result as $\textrm{IG}_i^j$. The contribution of the $i$-th input point to reconstructing the whole ECG cycle is measured by summing up the amplitudes of $\{\textrm{IG}_i^j|j=1,\cdots,L\}$, as $S_i=\sum_{j=1}^{L}|\textrm{IG}_i^j|$.

\begin{figure}[!ht]
\begin{center}
\includegraphics[width=0.6\linewidth]{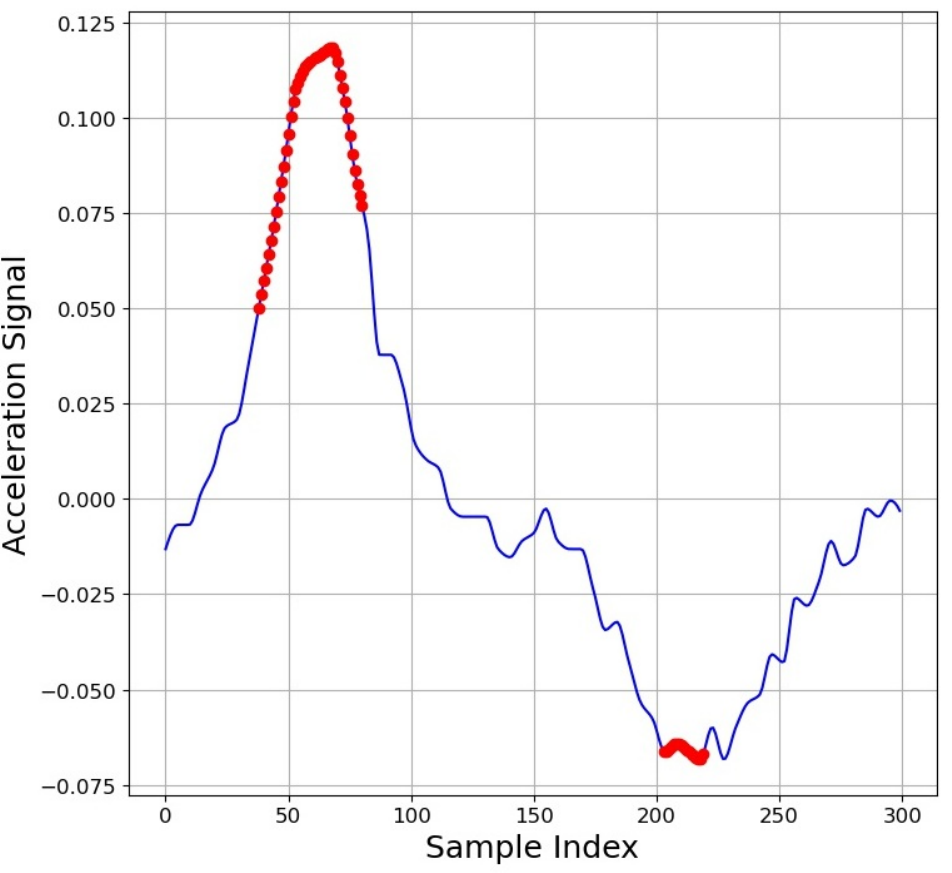}
\caption{Visualization of the model interpretation results about the acceleration signal. The sample points highlighted in red are the top 20\% ones that contribute most to the ECG reconstruction under exercise.}
\label{motion-Interp}
\end{center}
\end{figure}

By summing up $S_i$ over each input modality (PPG or acceleration signal), we found that the ratio of the contribution given by the acceleration signal in generating the output ECG is $27\%$. This is in line with our intuitive understanding. Despite the interference of motion, the activities of the heart are still the principal causing factor of PPG. Hence, a vast amount of the information for inferring ECG comes from PPG, while the acceleration signal only helps to reduce the motion artifacts of PPG. Let us further examine under what conditions the acceleration signal plays a more positive role in assisting ECG synthesis. In Fig.~\ref{motion-Interp}, we highlight the top $20\%$ sample points in the acceleration signal with the largest contribution. It is clear that the PPG2ECG network pays more attention to the auxiliary information when there are strong directional changes in acceleration. In such moments, the gap between the skin and the surface of the pulse oximeter changes rapidly, and this is the primary source of motion artifacts \cite{Motion-PPG}.

\subsection{Effects of Diagnosis-Oriented Training}
\label{sec:Effect-DOT}

We also evaluated the utility of reconstructed ECGs in diagnosing CVDs. The experiments were conducted on the MIMIC-III dataset where the testing waveforms are annotated with disease labels. The testing data cover five common CVDs, as listed in Section \ref{sec:DOT}. Reconstructed ECGs were fed to the CVD classifier described in Section \ref{sec:DOT}, and the classifier was trained using the original ECGs in the training subset of MIMIC-III. The diagnostic precision was evaluated at the cycle level. Diagnosis can also be made at the sequence level using majority voting and will exhibit higher precision. To set a more stringent criterion and eliminate the error-correcting effect of majority voting, we did not adopt the sequence-level precision for performance assessment. The disease label of each ECG cycle was predicted by picking the one with the highest probability. The diagnostic results were compared with the annotations provided by clinical staff, showing that $93\%$ of reconstructed ECGs are correctly classified. Table~\ref{F1} shows the disease-specific $F_1$ scores. For all kinds of diseases, the $F_1$ scores are over 0.95, indicating that the PPG2ECG network is able to infer the pathological ECG patterns from PPG.

\begin{table}[htbp]
\centering
\caption{Disease-specific diagnostic accuracy measured by $F_1$ score}\centering\label{F1}
    \begin{tabular}{c c c c c }
    \toprule
    CHF   & ST-MI & NST-MI   & HYPO   & CAD  \\ \hline
    0.96    & 0.96 & 0.95  &0.96 & 0.95 \\
    \bottomrule
    \end{tabular}
\end{table}

Ablation experiments were designed to demonstrate the effects of the diagnosis-oriented training (DOT) strategy. We re-trained the PPG2ECG network by removing $\mathcal{L}_F$ and $\mathcal{L}_C$ from (\ref{All-Loss}), only leaving the weighted $\ell_1$ norm. In Table~\ref{DOT-Abl}, we compare the performance of the networks trained with different approaches. After disabling DOT, the precision of CVD diagnosis on reconstructed ECGs drops to 0.90 from 0.93. It is worth mentioning that the training scheme also improves the fidelity of reconstructed signals, as verified by the comparison on $\rho$ and rRMSE in Table~\ref{DOT-Abl}. The feature loss and contrastive loss supervise the PPG2ECG network using the knowledge about the clustering structures of ECGs related to CVDs, so the network needs to represent the pathological changes of ECG with higher fidelity.

\begin{table}[htbp]
\centering
\caption{Effects of DOT measured on the proposed PPG2ECG network}\centering\label{DOT-Abl}
    \begin{tabular}{c c c c c c}
    \toprule
      Training Alg.  &$\bar{\rho} \uparrow$ &$\overline{\text{rRMSE}} \downarrow$ &$P_{NN}\uparrow$  &$P_{SVM}\uparrow$\\\hline
      With DOT   & 0.94 & 0.27  & 0.93 & 0.88 \\
      Without DOT & 0.93 & 0.29 & 0.90 & 0.83  \\
      \bottomrule
    \end{tabular}
\end{table}

We are curious if the benefits brought by DOT can extend to a different classifier that has not been used for regularizing ECG reconstruction, so experiments were also conducted in a model-agnostic manner by taking a multi-class support vector machine (SVM) as the classifier. The precisions attained by SVM on the ECGs reconstructed by the networks with and without DOT are 0.88 and 0.83, respectively (shown in the last column of Table~\ref{DOT-Abl}). We conjecture that the diagnostic cues captured by different classifiers have some overlaps, so the benefits brought by the regularization effect of one classifier are transferable to another unseen one.

The DOT algorithm is independent of the architecture of the ECG reconstruction model and can serve as a generic performance-boosting approach. In another experiment, we applied it to U-Net, and the network reinforced by our training scheme was compared with the one tested in Section \ref{sec:Perf} (see Table~\ref{U-Net-Train}). As in the above ablation experiment, performance gains in terms of the diagnostic precision of CVDs and the quality of output signals can be observed.

\begin{table}[htbp]
\centering
\caption{Effects of DOT measured on U-Net}\centering\label{U-Net-Train}
    \begin{tabular}{c c c c c c}
    \toprule
      Training Alg.  &$\bar{\rho} \uparrow$ &$\overline{\text{rRMSE}} \downarrow$ &$P_{NN}\uparrow$  &$P_{SVM}\uparrow$\\\hline
      With DOT   & 0.93 & 0.30  & 0.91 & 0.87 \\
      Without DOT & 0.93 & 0.32 & 0.86 & 0.80  \\
      \bottomrule
    \end{tabular}
\end{table}

A discriminative CVD classifier plays a strong effect in regularizing the PPG2ECG model. To examine the discriminating capability of the classifier, we further evaluated it on a more challenging dataset, PTB-XL \cite{PTB-XL}, which should be the largest publicly available ECG dataset to date. It contains 21,837 12-lead ECG waveforms measured from 18,885 patients over seven years, and each recording is of ten seconds. The dataset is a rich source of real-world cases, where sensing noise, drift, and other artifacts were all preserved in the raw waveforms. PTB-XL contains both normal and pathological ECG recordings, and the CVDs covered include conduction disturbance, myocardial infarction, hypertrophy, and ST/T change. Some diseases co-occur in a patient, making CVD prediction a multi-label classification problem, so the CVD prediction task on PTB-XL is more challenging than that on MIMIC-III. Accordingly, the number of output nodes in the classifier was set to five, each representing the probability of a specific label. We fed the 12-lead ECG waveforms to the classifier, so the input channel number was set to 12. To adapt the expressive power of the classifier to the diversity and scale of the dataset, we increase the kernel numbers of all convolutional layers by three times.

The training and testing sets were split using the method recommended by the creators of PTB-XL. Specifically, we used the waveforms in the tenth folder for testing and the remaining ones for training. Experiments were carried out on down-sampled versions of the ECG waveforms (with a sampling rate of 100Hz). We use the subject-centered metric to measure the performance of multi-label classification. Prediction results are obtained by comparing the estimated likelihood of each state with a decision threshold. For a given threshold, we calculate the precision and recall rates from the prediction results, and an $F_1$ score is computed for each subject. As the threshold varies from 0 to 1, the average $F_1$ score of all subjects observed at the optimal threshold is 0.71.

\subsection{Exploring CVD-Related Signs on PPG Using Model Interpretation}
\label{sec:CVD-PPG}

The CVD-related ECG abnormalities have been extensively studied, while the representations of CVDs on PPG are less understood. The pathological changes of the heart affect its pumping power and blood circulation, so PPG may reflect such abnormality. As shown earlier, the ECG signals reconstructed from PPG show encouraging performance in predicting CVDs, suggesting that PPG carries some diagnostic cues. Interpreting the mechanisms underlying data-driven models can offer medical practitioners complementary support and enrich the clinical knowledge base \cite{XAI}. Recall that we have constructed two models that establish the connections between PPG and ECG and between ECG and CVDs. The joint interpretation of these models has the potential to bridge the knowledge gap between PPG and CVDs.

Cascading the PPG2ECG network $G_{P\rightarrow E}(\cdot)$ and the ECG-based CVD classification network $\psi(\cdot)$ can produce a holistic classifier that directly predicts CVDs from PPG: $\Psi(\cdot)=\psi \circ G_{P\rightarrow E}(\cdot)$. As in Section \ref{sec:DOT}, we attribute the diagnostic result to each PPG point using the IG-based model interpretation \cite{IG}. The IG values tell which parts of PPG are most influential to the decision.

\begin{figure}[!ht]
\begin{center}
\includegraphics[width=1\linewidth]{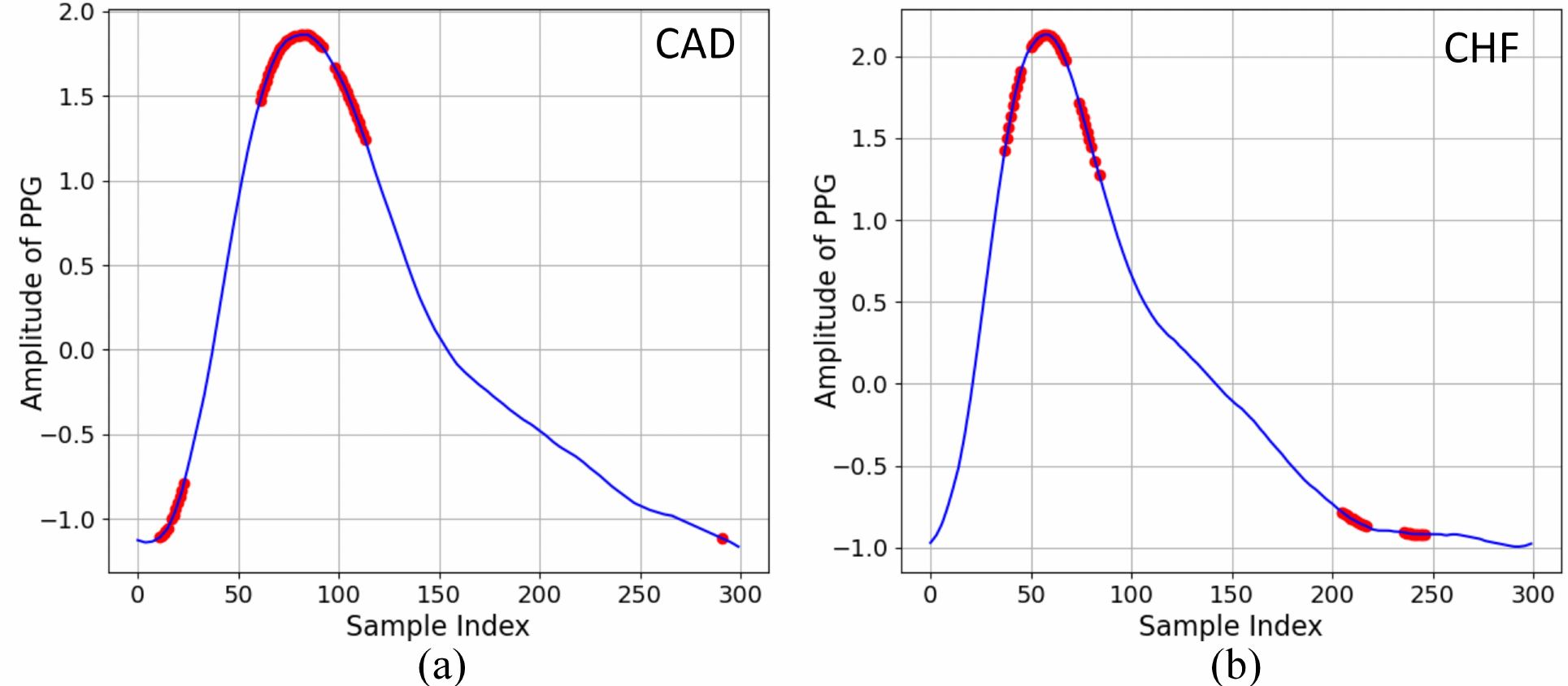}
\vspace{-0.5cm}
\caption{Visualization of the association between CVDs and PPG. (a) and (b) are the PPG cycles measured from two subjects diagnosed with CAD and CHF, respectively. For each disease, the sample points highlighted in red are the top 20\% most important ones for diagnosing the disease.}
\label{PPG-CVD}
\end{center}
\vspace{-0.5cm}
\end{figure}

In Fig.~\ref{PPG-CVD}, we show two PPG cycles measured from the subjects diagnosed with coronary artery disease (CAD) and congestive heart failure (CHF), respectively, and the red dots mark the top $20\%$ sample points contributing most to diagnosing the diseases. We find that for both diseases, the regions around the peak are more informative. It agrees with the finding that the increased rounding or triangulation and asymmetry are more likely to appear in the PPG waveforms of CVD patients \cite{PPG-CVD}. For CAD, the front of the ascending slope of PPG (corresponding to the moments when blood flows out of the heart) also receives high focus. CAD is caused by the plaque deposited on the inner walls of the arteries. According to the Moens-Korteweg equation \cite{MK-Equation}, the velocity of pulse wave partially depends on the elasticity of the arterial wall and the radius of the artery. The narrowing and increased stiffness of arteries affect the dynamics of blood flow. We conjecture that neural network learns to detect such changes from the increasing rate of blood volume. In the case of CHF, the heart becomes too weak to relax properly, resulting in reduced blood flow to the heart and the buildup of blood in other organs. From Fig.~\ref{PPG-CVD}(b), the sample points on the tail of PPG also contribute much to predicting CHF.

\subsection{Performance of Model Compression}
\label{sec:Exp-Comp}

In the model compression experiment, we replaced the four cascaded ENCs and DECs in the pre-trained PPG2ECG network with 2-depth recursive ENC and DEC to obtain a thinner student network. The kernel lengths of the three convolutional (or transposed convolutional) layers in the recursive modules are 13, 15, and 17. The student network approximates the distilled intermediate features and outputs of the pre-trained teacher network to minimize their performance gap. The training process only relies on a pre-trained network and input PPG signals without using the corresponding ECG signals. In this way, when adapting a pre-trained PPG2ECG network to lower-end devices with less memory and computational resources via distillation, there is no need to collect paired PPG and ECG signals, which alleviates the burden of data collection and reduces the information leakage in training data. In Table~\ref{Average-Compress}, we compare the model size, FLOPs, and average quality scores of the compressed network with other algorithms. Table~\ref{Compress} lists the quantitative performance of the full and compressed networks measured on each dataset. The compression cuts $67\%$ of parameters and $34\%$ of FLOPs. The compressed network only has 40K parameters and is about $5\sim273$ times smaller than the comparative ones. From Table~\ref{Compress}, the variations of average $\rho$ and rRMSE measured on all datasets after model compression are 0.01 and 0.02, respectively. Moreover, the loss of diagnostic precision is also maintained at a low level. The average $F_1$ scores of the neural-network-based CVD classifier measured on the waveforms reconstructed by the full and compressed networks are 0.96 and 0.94, respectively.

\begin{table}[h]
\centering
\caption{Comparison of the compressed PPG2ECG network with other algorithms}\centering\label{Average-Compress}
\begin{tabular}{ccccc}
\toprule
Algorithms & \#Para.(M)$\downarrow$ & FLOPs(M)$\downarrow$ &$\bar{\rho} \uparrow$ &$\overline{\text{rRMSE}} \downarrow$ \\\hline
DCT~\cite{DCT-IoT}    & 0.27   & 0.36 & 0.65  & 0.70   \\
XDJDL~\cite{SC-PPG2ECG}  & 5.67   & 60.21 & 0.74  & 0.55  \\
LC-XDJDL~\cite{SCLC-PPG2ECG}  & 5.90   & 64.72 & 0.75  & 0.53 \\
U-Net  & 0.22   & 55.99  & 0.86  & 0.44  \\
W-Net~\cite{W-Net}  & 2.37   & 35.97  &  0.87  & 0.45 \\
TANN~\cite{TANN}   & 10.98   & 518.79 & 0.88  & 0.40 \\ \hline
Full    & 0.12   & 28.32 & 0.88  & 0.40  \\
Compressed   & 0.04   & 18.63 & 0.87  & 0.42 \\
\bottomrule
\end{tabular}
\end{table}

\begin{table}[h]
\caption{Comparison between the full and compressed networks}\centering\label{Compress}
\resizebox{3.5in}{!}{
\begin{tabular}{cccccccc}
\toprule
\multirow{2}{*}{Datasets} & \multirow{2}{*}{Algorithms} & \multicolumn{3}{c}{$\rho \uparrow$} & \multicolumn{3}{c}{rRMSE $\downarrow$}                          \\ \cmidrule(r){3-5} \cmidrule(l){6-8}
 & & $\mu$ &med &$\sigma$ & $\mu$ &med &$\sigma$       \\ \hline
\multirow{2}{*}{MIMIC-III}
 & Full & 0.94 & 0.98 &0.13 &0.27 &0.22 &0.20\\
 & Compressed & 0.93 & 0.97 &0.14 &0.31 &0.25 &0.20\\ \hline
 \multirow{2}{*}{BIDMC}
 & Full & 0.90 & 0.97 &0.16 &0.35 &0.26 &0.26\\
 & Compressed & 0.89 & 0.95 &0.16 &0.39 &0.31 &0.25\\ \hline
 \multirow{2}{*}{\begin{tabular}[c]{@{}c@{}}Motion\\ PPG-ECG\end{tabular}}
 & Full & 0.81 & 0.83 &0.13 &0.57 &0.56 &0.16\\
 & Compressed & 0.80 & 0.83 &0.13 &0.57 &0.57 &0.16\\
 \bottomrule
\end{tabular}}
\end{table}

\subsection{Ablation Study on Neural Network Architectures}
\label{sec:Ablat}
We first examine the effects of the attention gates and the channel shuffle layers in ENCs and DECs. Three ablated networks were constructed by removing each and both kinds of modules from the PPG2ECG network. Table~\ref{Ablat} reports the average performance of each ablated network measured on the testing datasets. Compared with the original network, all the ablated networks show inferior performance. The average rRMSE between reconstructed and original ECG waveforms rises to 0.44 after discarding both modules. As mentioned above in Section \ref{sec:network}, the modules were designed to emphasize the informative parts of feature maps and maximize the fusion of multi-scale features, respectively. These mechanisms are beneficial to synthesizing fine-granular ECG structures. To control the parameter budget, the modules have quite few or no parameters. All the attention gates contain 432 parameters (approximately 0.3\% of the parameter amount of the PPG2ECG network), and the channel shuffle layers have no parameter.

\begin{table}[!htbp]
\centering
\caption{Comparison between the ablated and full networks}\centering\label{Ablat}
\begin{tabular}{cccc}
\toprule
Attention Gates & Channel Shuffle  &$\bar{\rho} \uparrow$ &$\overline{\text{rRMSE}} \downarrow$ \\\hline
$\times$ & $\times$  & 0.86  & 0.44  \\
$\surd$ & $\times$  & 0.86  & 0.43  \\
$\times$ & $\surd$  & 0.87  & 0.42  \\
$\surd$ & $\surd$  & 0.88  & 0.40   \\
\bottomrule
\end{tabular}
\end{table}

The kernel sizes of the convolutional layers in each ENC or DEC are set to different values to diversify the receptive fields in feature learning and ECG synthesis. We also conducted ablation experiments to verify the effectiveness of this design, where the kernel sizes were all set to nine (i.e., the average kernel size in the first ENC). Degradation in the qualities of reconstructed ECG cycles has been observed, where the average $\rho$ declines to 0.87 from 0.88, and the average rRMSE rises to 0.42 from 0.40.

\subsection{Leave-One-Out Mode Versus Personalized ECG Inference}
\label{sec:SubID-vs-Pers}

The results in Table~\ref{QualityScore} corroborate the generic correlations between PPG and ECG. It is also worth noting that the electrical and mechanical properties of the heart vary from individual to individual. Following the practice in another work \cite{SCLC-PPG2ECG}, we compared two training setups: the group mode and the leave-one-out mode, where the historical waveforms of a subject are included in and excluded from the training set, respectively. For a randomly selected subject in the BIDMC dataset, we measured the performance of the PPG2ECG networks trained using the two modes. For the group mode, the average $\rho$ measured on the subject is 0.92, and the value falls to 0.82 in the leave-one-out mode. This result is consistent with the observations that due to physiological variances, the rules learned from average population may not work best for all subjects \cite{SCLC-PPG2ECG}\cite{PGANs}. Patient-specific modeling (PSM) \cite{PSM} is a promising strategy for tackling this challenge. In light of individual variability and the difficulties of obtaining one-fit-all models, PSM advocates using the data of a target patient to develop individualized computational models for more accurate clinical outcomes. To meet the emerging trends of precision medicine and cardio-physiological digital twin, we also tested ECG inference in the PSM manner. Personalized ECG inference can be achieved by fine-tuning a pre-trained (in the leave-one-out mode) model using the historical data of an individual, which is more practical than training an individualized model from scratch. After two epochs of fine-tuning, the average $\rho$ rises to 0.97. Personalized fine-tuning benefits the subjects whose cardiovascular systems show some attributes rarely seen in the average population.

\section{Conclusions}
\label{sec:conclusion}
This paper has presented a computational approach for continuous ECG monitoring using optical sensing data. We have shown the feasibility of using a lightweight neural network to synthesize ECG waveforms from the blood volume variation signal measured by a PPG sensor. To facilitate CVD diagnosis, we have developed a task-aware training algorithm to ensure the precise representation of the clinically significant features of ECG. Compared with prior arts, the proposed algorithm demonstrates comparable or superior performance with fewer parameters. Our study also addresses the data analytic problems associated with PPG-based ECG synthesis and CVD diagnosis. We have leveraged model interpretation and causal analysis techniques to reveal the abnormal patterns of ECG and PPG related to CVDs, as well as the physiological and physical factors affecting PPG measurement. The insights gained from such exploration can support more transparent and reliable cardiac monitoring. We anticipate that the personalized PPG2ECG model may act as a building block of the cardio-physiological digital twin that enables personalized precision healthcare.

The current work has some limitations that need to be addressed in further studies. First, the current algorithm uses paired ECG and PPG waveforms in training, while synchronized or paired data are generally limited in practice, and a large volume of physiological data collected during routine monitoring are unpaired. It would be beneficial to investigate how to harness unpaired data to model the activities of the heart and overcome the scarcity of paired training data. Second, this work has not fully exploited the health information of a patient in ECG inference (e.g., age, weight, and medical history generally available in the health record). Such factors are known to be associated with certain CVDs and the pumping power of the heart. The ECG inference model conditioning on auxiliary health information may achieve more accurate and personalized cardiac monitoring.

\balance
\bibliographystyle{IEEEtran}
\bibliography{PPG2ECG}

\end{document}